%% file: main.tex
\documentclass[superscriptaddress,reprint,aps,nofootinbib,floatfix]{revtex4-2}

\usepackage{graphicx}
\usepackage{mathtools}
\usepackage{amssymb}
\usepackage[svgnames]{xcolor} 

\usepackage[colorlinks=True, linkcolor=SlateBlue,
            citecolor=Blue,urlcolor=BlueViolet]{hyperref}

\usepackage[capitalise]{cleveref} 

\usepackage{aas_macros}

\newcommand{\ip}[2]{\left(#1 \middle| #2\right)}
\newcommand{\mr}[1]{\mathrm{#1}}
\newcommand{\pd}{\partial}
\newcommand{\rf}[1]{\hat{\mathbf{#1}}}
\DeclareMathOperator{\re}{Re}
\newcommand{\imag}{\text{i}}
\newcommand{\diff}{\text{d}}
\newcommand{\vect}[1]{\boldsymbol{#1}}

\newcommand{\chieff}{\ensuremath{\chi_{\rm eff}}}
\newcommand{\chip}{\ensuremath{\chi_{\rm p}}}

\begin{document}

\author{Kevin A.\ Kuns}
\thanks{These two authors contributed equally}
\affiliation{LIGO Laboratory, California Institute of Technology, Pasadena, California 91125, USA}
\affiliation{LIGO Laboratory, Massachusetts Institute of Technology, Cambridge, Massachusetts 02139, USA}

\author{Hang Yu}
\thanks{These two authors contributed equally}

\affiliation{Theoretical Astrophysics 350-17, California Institute of Technology, Pasadena, California 91125, USA}


\author{Yanbei Chen}
\affiliation{Theoretical Astrophysics 350-17, California Institute of Technology, Pasadena, California 91125, USA}

\author{Rana X Adhikari}
\affiliation{Bridge Laboratory of Physics, California Institute of Technology, Pasadena, California 91125, USA}

\title{Astrophysics and cosmology with a decihertz gravitational-wave detector: TianGO}

\begin{abstract}
  We present the astrophysical science case for a space-based, decihertz gravitational-wave (GW) detector.
  We particularly highlight an ability to infer a source's sky location, both when combined with a network of ground-based detectors to form a long triangulation baseline, and by itself for the early warning of merger events.
  Such an accurate location measurement is the key for using GW signals as standard sirens for constraining the Hubble constant.
  This kind of detector also opens up the possibility to test type Ia supernovae progenitor hypotheses by constraining the merger rates of white dwarf binaries with both super- and sub-Chandrasekhar masses separately.
  We will discuss other scientific outcomes that can be delivered, including the constraint of structure formation in the early Universe, the search for intermediate-mass black holes, the precise determination of black hole spins,  the probe of binary systems' orbital eccentricity evolution, and the detection of tertiary masses around merging binaries.
\end{abstract}

\maketitle

\section{Introduction}
\label{sec:intro}

The coming decades will be an exciting time for gravitational-wave (GW) astronomy and astrophysics throughout the frequency band ranging from nano- to kilohertz. In the 10\,--\,10,000\,Hz band, detectors including Advanced LIGO (aLIGO)~\cite{lsc2015}, Advanced Virgo (aVirgo)~\cite{Acernese2015}, and KAGRA~\cite{Akutsu2018} are steadily improving towards their sensitivity goals.
Meanwhile, various upgrades to current facilities have been proposed, including the incremental A+ upgrade~\cite{lsc2019} and the Voyager design which aims to reach the limits of the current infrastructure~\cite{VoyagerWhite}.
In the long run, third generation detectors including the Einstein Telescope~\cite{Hild2010, Sathyaprakash2012} and Cosmic Explorer~\cite{Evans2017} are expected to push the audio-band reach of GW astronomy out to cosmological distances.
In the millihertz band, space-borne laser interferometers such as LISA~\cite{Amaro-Seoane2017} and TianQin~\cite{Luo2016, Wang2019} would give us exquisitely sensitive probes fo many astrophysical signals -- both are planned to be launched around 2035.
At even lower frequencies, pulsar timing arrays are becoming evermore sensitive with more pulsars being added to the network~\cite{Parkes:2015, Verbiest2016, NANOGrav2018}.
Nonetheless, gaps still exist between these missions. This especially limits our ability to have a coherent, multi-band coverage of the same source; even a relatively massive 30\,$M_\odot$-30\,$M_\odot$ black hole (BH) binary at 0.01\,Hz (where LISA is most sensitive) will not enter a ground-based detector's sensitive band until 20 years later. 

Therefore, we propose a space-based detector, TianGO, which is sensitive in the 0.01\,--\,10\,Hz band and which fills the gap between LISA and the ground-based detectors~\cite{InstrumentPaper}. A possible advanced TianGO (aTianGO) would have 10 times better sensitivity, but is not discussed further here. In this paper we expand on the pioneering work of Ref.~\cite{Mandel2018} and explore the scientific promise of TianGO. Our work also sheds light on other decihertz concepts~\cite{Sato2017,Harry2006}.

\begin{figure}
  \centering
  \includegraphics[width=\columnwidth]{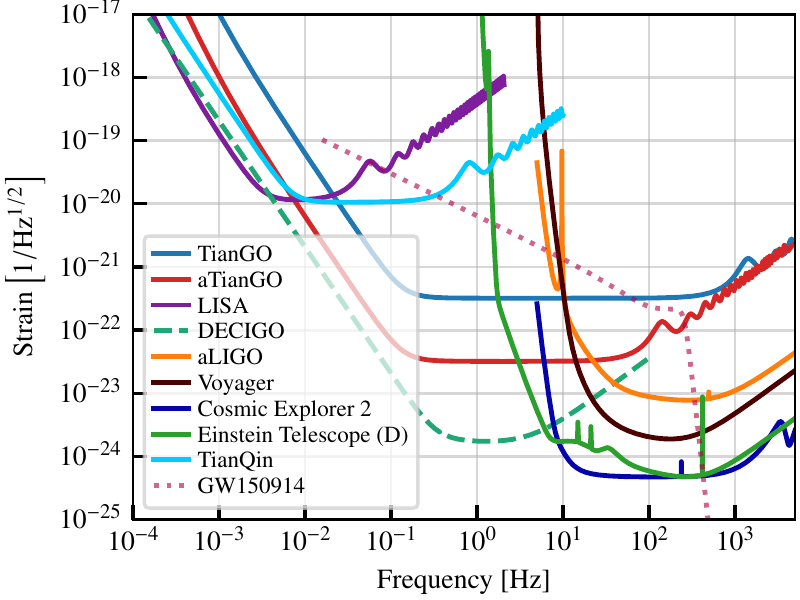}
  \caption{Sensitivities of future ground-based and space gravitational wave detectors. The sensitivities are averaged over sky location and polarization. The LISA curve includes two $60^\circ$ interferometers and the ET curve includes three $60^\circ$ interferometers. The curve labeled ``GW150914'' is $2\sqrt{f}h$, where $h$ is the waveform of the first gravitational wave detected~\cite{Abbott2016} starting five years before merger.
  }
  \label{fig:network}
\end{figure}

\begin{figure}
  \centering
  \includegraphics[width=\columnwidth]{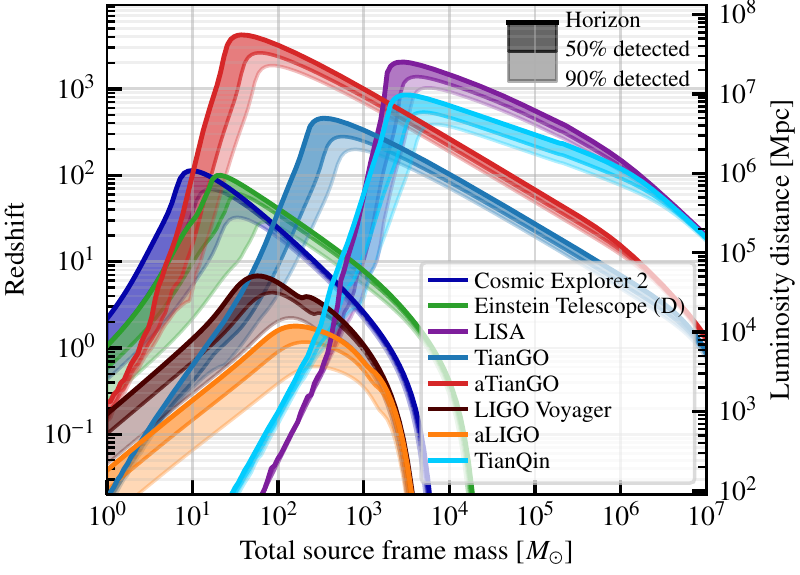}
  \caption{Horizons for equal mass compact binaries oriented face on for the detectors shown in \cref{fig:network}. The maximum detectable distance, defined as the distance at which a source has an SNR of 8 in a given detector, is computed for 48 source locations uniformly tiling the sky. The horizon is the maximum distance at which the best source is detected, 50\% of these sources are detected within the dark shaded band, and 90\% of the sources are detected within the light shaded band. If a source stays in a space detector's sensitivity band for more than 5 years, the 5 year portion of the system's evolution that gives the best SNR in each detector is used.
  }
  \label{fig:horizons}
\end{figure}

\cref{fig:network} shows the sensitivity of TianGO and other major detectors. For the rest of the paper, unless otherwise stated, the ground-based detectors are assumed to have the Voyager design sensitivity~\cite{VoyagerWhite} and the ground-based network consists of the three LIGO detectors at Hanford (H), Livingston (L), USA, and Aundha, India (A); Virgo (V) in Italy; and KAGRA (K) in Japan. The corresponding detection horizons for compact binary sources of different total mass are shown in \cref{fig:horizons}. For stellar-mass compact objects such as neutron stars (NSs) and BHs, TianGO has a comparable range as the ground-based detectors. Moreover, even a relatively light NS binary starting at 0.12\,Hz, where TianGO is most sensitive, will evolve into the ground-based detectors' band and merge within 5 years. This facilitates a multi-band coverage of astrophysical sources.

\begin{table*}[tb]
 \caption{Summary of TianGO science cases}
 \label{tab:summary}
 \begin{ruledtabular}
 \begin{tabular}{ccccc}
 Section                                &  Scientific Objective                   			& Target           				& Information to extract 				& Key references  \\
  \hline
 \ref{sec:cosmography}    & Cosmography.  			& Binary BHs   				& Sky location                			&  \cite{Schutz1986, Sathyaprakash2010,Kyutoku2017}  \\
 \ref{sec:early-warning}      & Multi-messenger astrophysics; NS physics.   & Binary NSs   				& Sky location                			& \cite{Burns2019} \\
 \ref{sec:IMBH}                   & Structure formation; IMBHs.   	 		& Binaries involving IMBHs 	& Source population 				& \cite{Hughes2002, Sesana2007, Sesana2009} \\
 \ref{sec:typeIa}                  & Type-Ia SNe progenitors.           			& Binary WDs   			& Source population      			& \cite{Marsh11, Maoz2012, Maoz2014} \\
 \ref{sec:WDtide}		   & WD physics.                            			& Binary WDs                            & Tidal dephasing                      	& \cite{Fuller2012, Burkart2013, Yu2020} \\
 \ref{sec:BHspin}                & Formation of binary BHs; Stellar evolution.	& Binary BHs                             & Aligned and precession spin           & \cite{Rodriguez2016, Fuller2019b} \\
 \ref{sec:eccOrb}                & Formation of binary BHs.                                & Binary BHs                             & Orbital eccentricity                          & \cite{Barack2004, Chen2017} \\
 \ref{sec:third_mass}          & Environment around BHs.                              & Binary BHs                             & Phase modulation			         & \cite{Randall2019} 
 \end{tabular}
 \end{ruledtabular}
 \end{table*}

In particular, by placing TianGO in an orbit from between a 5 and 170\,s light travel time from the Earth, the localization of astrophysical sources is significantly improved over that possible with a ground-based network alone: when combined with the ground-based network, this long baseline allows a combined TianGO-ground-based network to increase the angular resolution by a factor of ${\sim}50$ over that of the ground-based network alone. This exquisite ability to localize sources enables this hybrid network to do precision cosmography. Furthermore, since a binary of two NSs or of a NS and BH will stay in TianGO's sensitivity band for several years, TianGO will provide an early warning for the ground-based GW detectors and the electromagnetic telescopes.


Meanwhile, there are astrophysical sources that are particularly well suited to be studied by a decihertz detector like TianGO. For example, intermediate-mass black holes (IMBHs) are one of such examples. TianGO is sensitive to the mergers of both a binary of IMBHs and an IMBH with a stellar-mass compact companion. Consequently, TianGO will be the ideal detector to either solidly confirm the existence of IMBHs with a positive detection or strongly disfavor their existence with a null-detection. Meanwhile, mass transfer starts at $\sim$30\,mHz for a typical white dwarf (WD) binary.
This frequency will be higher for even more massive, super-Chandrasekhar WD binaries. As LISA's sensitivity starts to degrade above 10\,mHz, TianGO will be the most sensitive instrument to study the interactions between double WDs near the end of their binary evolution, which may be the progenitors of type-Ia supernovae.  Lastly, if a system is formed with a high initial eccentricity, TianGO will be able to capture the evolution history of the eccentricity, which will in turn reveal the system's formation channel.

We summarize the major science targets we will be considering in \cref{tab:summary} and also in the text as follows.
We discuss the precision with which BBHs can be localized with a hybrid network and the application to cosmography in \cref{sec:cosmography}.
We then examine TianGO's ability to localize coalescing binary NSs and to serve as an early warning for ground-based and EM telescopes, the most crucial component for multi-messenger astrophysics, in \cref{sec:early-warning}.
In \cref{sec:IMBH} we discuss the possibility of using TianGO to distinguish the cosmological structure formation scenarios and to search for the existence of IMBHs.
This is followed by our study of the progenitor problem of type Ia supernovae in \cref{sec:typeIa}.
We then discuss the detectability of tidal interactions in binary WDs with TianGO in \cref{sec:WDtide}.
In \cref{sec:BHspin} we analyze TianGO's ability to accurately determine both the effective and the precession spin, and how we may use it to constrain the formation channels of stellar-mass BH binaries as well as the efficiency of angular momentum transfer in the progenitor stars.
In \cref{sec:eccOrb} we explore TianGO's capability of measuring the orbital eccentricity evolution.
In \cref{sec:third_mass} we discuss TianGO's ability to directly probe the existence of tertiary masses around merging binaries.

\section{Gravitational-wave Cosmography}
\label{sec:cosmography}
The Hubble constant, $H_0$, quantifies the current expansion rate of the universe, and is one of the most fundamental parameters of the standard $\Lambda$CDM cosmological model, yet the two traditional methods of measuring it disagree at the $4.4\sigma$ level~\cite{Riess2019}.
The first method relies on the physics of the early universe and our understanding of cosmology to fit observations of the CMB to a cosmological model~\cite{Ade2016}.
The second, local measurement, relies on our understanding of astrophysics to calibrate a cosmic distance ladder. This ladder relates the redshifts of observed sources to their luminosity distances~\cite{Riess1998,Perlmutter_1999,Riess2019}. Gravitational wave astronomy adds a third method of determining $H_0$ and the prospect of resolving this tension~\cite{Schutz1986,Holz2005,Cutler2009,Sathyaprakash2010,Kyutoku2017}, a task for which a combined TianGO-ground-based network is particularly well suited.

To obtain the redshift-distance relationship necessary to determine $H_0$, the local measurement first determines the redshift of a galaxy. The luminosity distance cannot be measured directly, however, and relies on the calibration of a cosmic distance ladder to provide ``standard candles.'' On the other-hand, the luminosity distance is measured directly from a GW observation requiring no calibration and relying only on the assumption that general relativity describes the source. This makes gravitational waves ideal ``standard sirens.'' If the host galaxy of a gravitational wave source is identified, optical telescopes can measure the redshift.\footnote{The GW standard sirens can also be used to independently calibrate the EM standard candles forming the cosmic distance ladder~\cite{Gupta2019}.} In this way, both the redshift and the distance are measured directly. The BNS GW170817 was the first GW source observed by both gravitational and electromagnetic observatories~\cite{GW170817a}. Since the gravitational wave signal was accompanied by an optical counterpart, the host galaxy was identified and the first direct measurement of $H_0$ using this method was made~\cite{GW170817c}.

\begin{figure}
  \centering
  \includegraphics[width=\columnwidth]{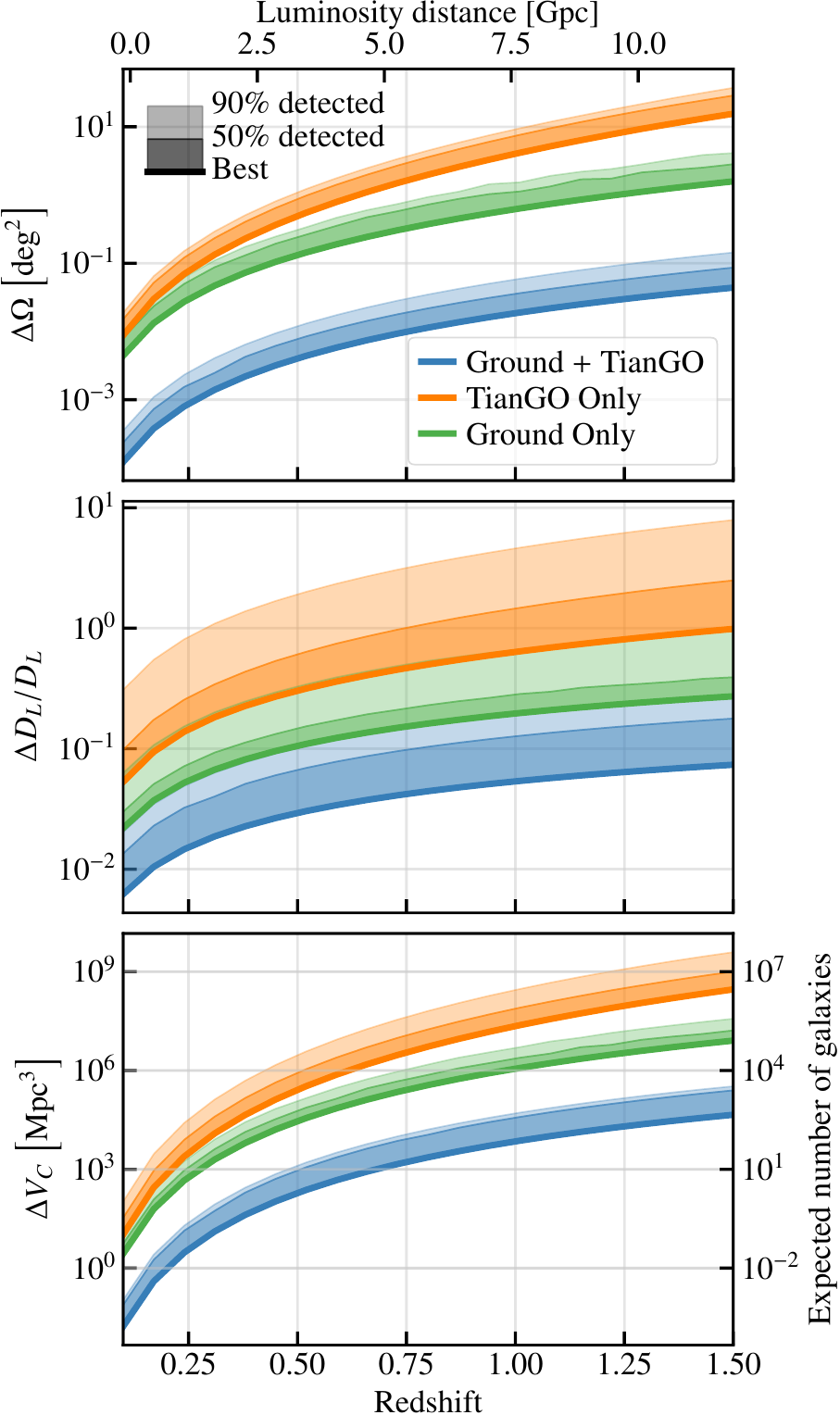}
\caption{Sky localization, luminosity distance, and volume localization precision as a function of redshift for a binary black hole system with $\mathcal{M}_c=25\,M_\odot$, $q=1.05$, and an inclination $\iota = 30^\circ$ and TianGO in a $5^\circ$ Earth trailing orbit. A notional density of $0.01\,\mr{galaxies}/\mr{Mpc}^3$ is used to convert $\Delta V_C$ to the expected number of galaxies. \cref{fig:sky-localization-0deg} shows the same for face-on binaries.}
\label{fig:sky-localization-30deg}
\end{figure}

Identifying the host galaxy to make these measurements requires precise sky localization from the GW detector network. This ability is greatly enhanced when TianGO is added to a network of ground-based detectors. TianGO will either be in a Earth-trailing orbit of up to $20^\circ$ or an orbit at the L2 Lagrange point~\cite{InstrumentPaper} thereby adding a baseline of between $1.5\times 10^6\, \mr{km} = 235 R_\oplus$ and $5.2\times 10^7\, \mr{km}=8.2\times10^3 R_\oplus$ to the network, where $R_\oplus$ is the radius of the Earth. Since the same source will be observed by both TianGO and the ground-based network, the timing accuracy formed by this large baseline significantly improves the sky localization ability over that of the ground-based alone, as is illustrated in \cref{fig:sky-localization-30deg,fig:sky-localization-0deg} and \cref{tab:sky_vs_network}.

The top panel of \cref{fig:sky-localization-30deg} shows the angular resolution $\Delta\Omega$ as a function of redshift as determined from the network of ground-based detectors alone, TianGO alone, and the combined network of the ground-based and TianGO in a $5^\circ$~Earth-trailing orbit. The source is a BBH with $\mathcal{M}_c=25\,M_\odot$, $q=1.05$, and $\iota=30^\circ$. (The probability of detecting binaries with a given inclination peaks around $\iota=30^\circ$~\cite{Schutz2011}. The same figure for $\iota=0^\circ$ is shown in \cref{fig:sky-localization-0deg}.) The extra long baseline formed by TianGO and the ground-based network improves the angular uncertainty by a factor of $\sim 50$.

The middle panel of \cref{fig:sky-localization-30deg} shows the fractional uncertainty $\Delta D_L/D_L$ in measuring the luminosity distance. Note that the inference accuracy for the ground-based network is limited by the distance-inclination degeneracy. (This is especially true for face-on sources as can be seen by comparing \cref{fig:sky-localization-30deg,fig:sky-localization-0deg}.) TianGO breaks this degeneracy due to the time-dependent antenna pattern caused by its tumbling orbit. The combined TianGO-ground-based uncertainty is thus significantly better that of the ground-based alone.

The bottom panel of \cref{fig:sky-localization-30deg} shows the uncertainty in comoving volume localization $\Delta V_C$.\footnote{The Planck 2015 cosmology is assumed~\cite{Ade2016}.} If an optical counterpart is not observed, or does not exist as is likely for most of the sources for which the TianGO-ground-based network will be sensitive, the GW detector network must localize the host to a single galaxy. To estimate the number of galaxies contained in a comoving volume $\Delta V_C$, the value of $0.01\,\mr{galaxies}/\mr{Mpc}^3$ is assumed. The combined network can localize a source to a single galaxy up to a redshift of $z\sim 0.5$ for the best face-on sources, and to $z\sim 0.35$ for the median sources at $\iota = 30^\circ$.

Even if the host galaxy cannot be uniquely identified, galaxy catalogs can be used to make a statistical inference about the location of the source~\cite{Schutz1986,DelPozzo2012,MacLeod2008,Nair2018,Arabsalmani2013}. This method has been used to reanalyze the measurement from GW170817 to infer $H_0$ without the unique galaxy determination provided by the observation of the optical counterpart~\cite{Fishbach2019} and has been used to improve the original analysis of Ref.~\cite{GW170817c} with further observations of BBHs without optical counterparts~\cite{O2H0}. Future work will quantify the extent to which the TianGO-ground-based network's exquisite sky localization can improve the reach of these methods.

\begin{table}
 \caption{Comparison of sky localization for different networks of detectors. The neutron star system is a binary with $\mathcal{M}_c = 1.2\,M_\odot$, $q=1.05$, and $D_L=50\,\mr{Mpc}$. The black hole system is a binary with $\mathcal{M}_c = 25\,M_\odot$, $q=1.05$, and $D_L=600\,\mr{Mpc}$. The best and median sources are given in $\mr{deg}^2$. The ground-based detectors have the Voyager design sensitivity.}
 \label{tab:sky_vs_network}
 \begin{ruledtabular}
\begin{tabular}{c|cc|cc}
 & \multicolumn{2}{c|}{Neutron Star} & \multicolumn{2}{c}{Black Hole} \\
Network & Best & Median & Best & Median \\
\hline
\input{LocalizationTable.tex}
\end{tabular}

 \end{ruledtabular}
 \end{table}

\section{Early Warning of Binary Neutron Star Coalescence}
\label{sec:early-warning}

The joint detection of a coalescing binary NS in GW~\cite{GW170817a} and $\gamma$-ray~\cite{Goldstein2017}, and the follow-up observation of the post-merger kilonova in electromagnetic radiations~\cite{GW170817b} heralds the beginning of an exciting era of multi-messenger astronomy. While the first detection has provided some valuable insights on the nature of short $\gamma$-ray bursts and kilonovae, significantly more are expected to come from future multi-messenger observations~\cite{Burns2019}.
The success of such a joint observation relies critically on the GW observatories to produce an accurate sky map of the source's location in a timely manner, and TianGO is an ideal instrument to perform the early warning and localization of coalescing compact binaries.
As a typical NS binary will stay in TianGO's band for a few years before the final merger, the Doppler phase shift and time-dependent antenna patterns due to TianGO's orbital motion enables it to localize the source by itself with high accuracy.

\begin{figure}
  \centering
  \includegraphics[width=\columnwidth]{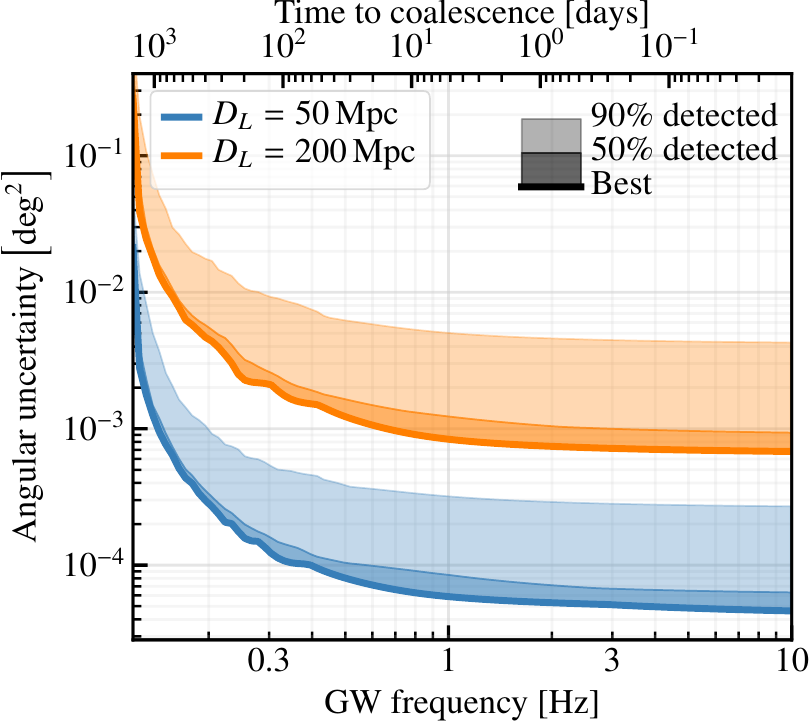}
  \caption{Angular uncertainty as determined by TianGO alone for a face-on BNS (at 12 source locations uniformly tiling the sky) with $(M_1,\, M_2)=(1.4\,M_\odot,\,1.35\,M_\odot)$ as a function of GW frequency or time to coalescence.
  }
  \label{fig:BNSloc}
\end{figure}

This is illustrated in detail in \cref{fig:BNSloc} and \cref{tab:sky_vs_network}. \cref{fig:BNSloc} shows the {\it cumulative angular uncertainty} for a typical NS binary with $(M_1,\, M_2)$=$(1.4\,M_\odot,\,1.35\,M_\odot)$. More specifically, on the bottom of the frame we show the GW frequency up to which we integrate the data, and on the top of the frame we show the corresponding time to the final merger, given by
\begin{equation}
t_{\rm m}(f) = 5.4 \left(\frac{\mathcal{M}_c}{1.2\,M_\odot}\right)^{-5/3}\left(\frac{f}{1\,{\rm Hz}}\right)^{-8/3}\,{\rm days}.
\label{eq:t_merger}
\end{equation}
We assume that the source has a face-on orientation, yet we vary its right ascension and declination to cover the entire sky. Two representative distances, $D_L=50\,{\rm Mpc}$ and $D_L=100\,{\rm Mpc}$, are shown in the plot. With TianGO alone, we can localize the majority of sources to within a few$\times10^{-3}$ $\mr{deg}^2$ approximately 10 days before the final merger.  This provides sufficient time for the GW network to process the data and inform the electromagnetic observatories to prepare the telescopes for the final merger. 

Furthermore, the localization accuracy for NS binaries obtained by TianGO alone is in fact nearly 100 times better than a network of 5 ground-based detectors each with Voyager's designed sensitivity (LHVKA; see Table~\ref{tab:sky_vs_network}), and is much smaller than the typical field of view of an optical telescope of $\mathcal{O}(1)$ square-degree.

In addition to post-merger emissions, TianGO also significantly enhances the possibility of capturing the potential precursor emissions during the inspiral phase (see, e.g., Section 2.2 of Ref.~\cite{Fernandez2016}). One example is the energy release due to shattering of the NS crust~\cite{Tsang2012}, which is suspected to be the source of short-$\gamma$-ray burst precursors~\cite{Troja2010}. The timing when the precursor happens is directly related to the equation of state of materials near the crust-core interface. Additionally, if at least one of the NS is highly magnetized, the orbital motion during the inspiral may also trigger an electron-positron pair fireball that will likely emerge in hard X-ray/gamma ray~\cite{Metzger2016}. In the radio band, the magnetospheric interaction may also extract the orbital energy and give rise to a short burst of coherent radio emission~\cite{Hansen2001}. Such an emission could be a mechanism leading to fast radio bursts~\cite{Thornton2013}. With TianGO's ability to accurately pinpoint a source days prior to the merger, one can unambiguously associate a precursor emission at the right time and location to coalescing binary NSs. 

\section{Cosmological structure formation and intermediate-mass black holes}
\label{sec:IMBH}
Massive BHs reside in the center of most local galaxies. Despite the fact that the mass of the central BH is only ${\sim} 0.1\%$ of the total mass of the host galaxy, surprisingly clear correlations between the massive BH's mass and the properties of the host galaxy have been observed (e.g., Ref.~\cite{Beifiori2012}). This thus suggests a co-evolution of the massive BH and its host galaxy~\cite{Kormendy2013}, which is further sensitive to the seed from which the massive BHs grow (see Ref.~\cite{Volonteri2010} for a review). Broadly speaking, a massive BH may grow from either a ``heavy seed'' with mass $\sim 10^4-10^6\,M_\odot$ at a relatively late cosmic time of $z{\sim} 5-10$, or from a ``light seed'' with mass ${\simeq} 100-600\,M_\odot$ at an earlier time of $z{\simeq} 20$. Those ``light seeds'' may be generated from the collapse of Pop III stars~\cite{Madau2001} and they may merge with each other in the early Universe~\cite{Volonteri2003}. 

The characteristic frequency of such a merger is given by the system's quasi-normal mode frequency. For a Schwarzschild BH, the fundamental, axially symmetric, quadrupolar mode oscillates at a frequency of~\cite{Kokkotas1999},
\begin{equation}
f^{\rm (det)}_{\rm QNM} \simeq 1.21\left(\frac{10}{1+z}\right)\left(\frac{10^3\,M_\odot}{M_1 + M_2}\right)\, {\rm Hz}. 
\end{equation}
We have used the superscript ``(det)'' to represent quantities measured in the detector-frame. 
While it is a frequency too low for ground-based detectors and too high for LISA, it falls right into TianGO's most sensitive band. Indeed, as shown in Figure~\ref{fig:horizons}, TianGO is especially sensitive to systems with masses in the range of $100{-}1000\,M_\odot$ and can detect them up to a redshift of $z{\sim} 100$. Consequently, if massive BHs grow from light seeds, TianGO will be able to map out the entire growth history throughout the Universe. On the other hand, a null detection of such mergers by TianGO can then rule out the ``light seed'' scenario. It will also constrain our models of Pop III stars that will be otherwise challenging to detect even with the \textit{James Webb Space Telescope}~\cite{Rydberg2013}. In either case, TianGO will provide indispensable insights in our understanding of cosmological structure formation (see also Refs.~\cite{Hughes2002, Sesana2007, Sesana2009} for relevant discussions for LISA and the third-generation ground-based GW observatories).

Meanwhile, those seed BHs that failed to grow into massive and supermassive BHs may be left to become IMBHs in the local Universe~\cite{Koliopanos2017, Mezuca2017}. While a few IMBH candidates have been reported (see, e.g., ~\cite{Filippenko2003, Farrell2009, Kiziltan2017}), a solid confirmation is still lacking from electromagnetic observation. This makes the potential GW detection of an IMBH particularly exciting. In addition to the merger of two IMBHs (similar to the mergers of light BH seeds discussed above), another potential GW source involving an IMBH is the intermediate-mass-ratio inspirals (IMRIs): a stellar mass object (BH, NS, or WD) merges with an IMBH.  IMRIs may be found in the dense cores of globular clusters~\cite{Haster2016a, Haster2016b}. 

TianGO will detect a typical IMRI source with $(M_1, M_2){=}(1000\,M_\odot,\, 10\,M_\odot)$ at $z{=}1$ with an SNR of 10 after averaging over both orientation and sky location. If the event rate for such a merger is about 1 per Gpc$^3$ per year as argued in Ref.~\cite{Mandel2018}, we would be able to detect nearly 1000 IMRI mergers over a 5-year observation period. The numerous detections would thus allow us to both place constraints on the dynamics in globular clusters and perform potential tests of general relativity in a way similar to those using the extreme-mass-ratio inspirals~\cite{Hughes2006}.

\section{Binary White Dwarves as Progenitors of Type Ia Supernovae }
\label{sec:typeIa}
Type Ia supernovae are one of the most powerful family of standard candles for determining the cosmological distance~\cite{Howell2011} and they have led to the discovery of the accelerating expansion of the Universe~\cite{Riess1998}. However, the identity of their progenitors remains an unresolved problem in modern astrophysics despite decades of research. Among all possibilities, the merger of two WDs (also known as the double-degenerate progenitor) is an increasingly favored formation channel, yet it is still unclear if the system's total mass exceeding the Chandrasekhar limit is a necessary condition for a supernova explosion (for recent reviews, see Refs.~\cite{Maoz2012, Maoz2014}). In this section we show how TianGO can help to improve our understanding of the problem (see Ref.~\cite{Marsh11} for a similar discussion for LISA).

\begin{figure}[tb]
  \centering
  \includegraphics[width=\columnwidth]{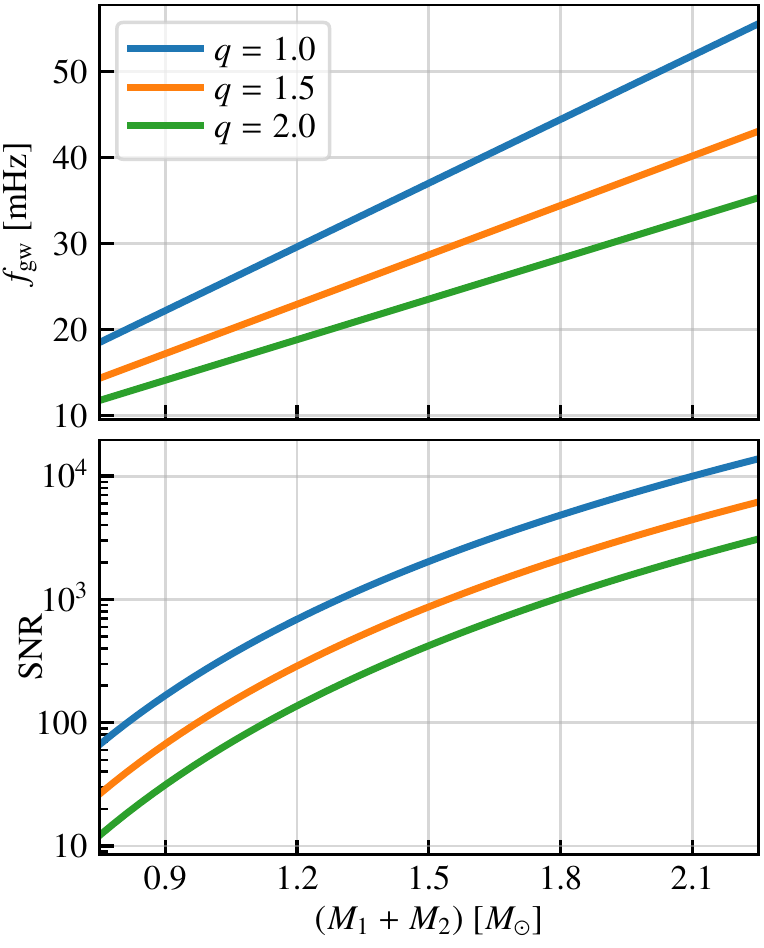}
\caption{Upper panel: the GW frequency $f_{\rm gw}$ for WD binaries with different total masses $(M_1+M_2)$ and mass ratios $q\equiv M_1/M_2\geq 1$ at the onset of Roche-lobe overflow. Bottom panel: angle-averaged SNR seen by TianGO, assuming a source distance of 10\,kpc and an observation period of 5 years.}
\label{fig:fSnrAtRoche}
\end{figure}

The key is that TianGO is capable of individually resolve essentially \emph{all} the Galactic WD binaries when they are close to starting or have just started mass transfer. This is illustrated in Figure~\ref{fig:fSnrAtRoche}. In the upper panel, we show the GW frequency for WD binaries at the onset of the Roche-lobe overflow. Here we assume a simple mass-radius relation for WDs as 
\begin{equation}
R_{\rm wd}(M_{\rm wd}) = 10^9 \left(\frac{M_{\rm wd}}{0.6\,M_\odot}\right)^{-1/3}\,{\rm cm},
\label{eq:RvsM}
\end{equation}
and we find the orbital separation such that the donor star's radius is equal to the volume-equivalent radius of its Roche lobe~\cite{Eggleton1983}. For such systems, the SNR (averaging over both sky location and source orientation) seen by TianGO over a 5-year observation period is shown in the lower panel. The source distance is fixed at 10\,kpc. TianGO thus allows us to construct thorough statistics on the WD population which can further be used to calibrate theoretical population synthesis models (e.g., Refs.~\cite{Hurley2002, Toonen2012}). Then, comparing the merger rate of double WDs predicted in the model to the observed type Ia supernovae rate allows a test of the double-degenerate progenitor hypothesis. 

Specifically, for a population of WDs driven by GW radiation only, the number density per orbital separation $n(a)$ should scale with the orbital separation $a$ as
\begin{equation}
n(a) \propto 
    \begin{cases}
        a^{3} &\text{ for }\alpha \geq -1, \\
        a^{\alpha + 4} &\text{ for }\alpha < -1,
    \end{cases}
\end{equation}
where $\alpha$ is the power-law index of the population's initial separation distribution. This scaling is valid for binaries with a current separation of $a\ll0.01\,\mr{AU}$ and prior to Roche-lobe overflow. Once we determine the constant of proportionality with TianGO, we can then predict the merger rate as $n(a) {\rm d} a/{\rm d}t$~\cite{Maoz2012}. 

While LISA is expected to detect a similar number of WD binaries as TianGO, there are nonetheless unique advantages of TianGO in constraining the binary WD population. Note that a WD binary in LISA's more sensitive band of 1-20\,mHz will evolve in frequency by so little over a $\sim$ 5-year observation that it either is unresolvable or can only be used to measure the system's chirp mass. In the case of the type Ia supernovae progenitor problem, however, it is the system's total mass and mass ratio that are of interest. TianGO, on the other hand, is more sensitive to systems at higher frequencies ($\gtrsim 20\,{\rm mHz}$) and therefore will see a greater amount of frequency evolution. Moreover, those systems will experience a stronger tidal effect which depends on the masses in a different way than the chirp mass, allowing for a determination of the component masses (see \cref{sec:WDtide} for more details). Consequently, with TianGO we can determine the distributions for double WD systems with different total masses. This is critical for examining the possibility of sub-Chandrasekhar progenitors~\cite{Sim2010, vanKerkwijk2010, Woosley2011, Polin2019}. 

At the same time, TianGO will also be able to identify the deviation of the power-law distribution for different WD binaries due to the onset of mass transfer. The stability of the mass transfer is a complicated problem that depends on factors like the system's mass ratio, the nature of the accretion, and the efficiency of tidal coupling~\cite{Marsh2004, Gokhale2007, Dan2011, Kremer2015, Kremer2017}. TianGO will provide insights on this problem by both locating the cutoffs in the distribution that marks the onset of unstable mass transfer, and measuring directly the waveforms of the surviving systems that may evolve into AM CVn stars~\cite{Nelemans2005}. TianGO also has the potential of resolving the current tension between the observed spatial density of AM CVn stars and that predicted by population synthesis models~\cite{Roelofs2007}.  

\section{Detecting White Dwarf Tidal Interactions}
\label{sec:WDtide}
When a WD binary's orbit decays due to GW radiation, tidal interaction starts to play an increasingly significant role in its evolution. In this section we discuss the prospects of detecting tides in WDs with TianGO.

The tidal response of a fluid can be decomposed into an equilibrium component and a dynamical component. In the equilibrium tide, the fluid distribution follows the gravitational equipotential instantaneously.  In most situations, this already captures the large-scale distortion of the star.  The dynamical tide, on the other hand, accounts for the star's dynamical response to the tidal forcing and represents the excitation of waves. Whereas for NSs in coalescing binaries the equilibrium component dominates the tidal interaction~\cite{Flanagan2008, Hinderer2010, lsc2018c},  for WDs in binaries, it is the dynamical tide that has the most significant effect. 

As shown in Refs.~\cite{Piro2011, Fuller2012, Fuller2012b, Burkart2013, Piro2019, Yu2020}, when a WD binary enters TianGO's band, the dynamical tide can keep the WD's spin nearly synchronized with the orbit. Consequently,
\footnote{Here we ignore the rotational modification of the WD structure, as the Coriolis force only mildly modifies the tidal dissipation in subsynchronously rotating WDs~\cite{Fuller2014}.}
\begin{equation}
\dot{\Omega}_{\rm s,1}\simeq\dot{\Omega}_{\rm s, 2}\simeq \dot{\Omega}_{\rm orb},\label{eq:dOmegaSpinOrb}
\end{equation}
where $\Omega_{\rm s, 1(2)}$ is the angular spin velocity of mass 1 (2). In terms of energy, we have
\begin{equation}
\frac{\dot{E}_{\rm tide1 (2)}}{\dot{E}_{\rm pp}}\simeq\frac{3}{2}\frac{I_{1(2)} \Omega_{\rm orb}^2}{E_{\rm orb}}  \propto f^{4/3}.\label{eq:dEtide}
\end{equation}
Here $\dot{E}_{\rm tide1 (2)}$ is the amount of  energy transferred per unit time from the orbit to the interior of mass 1(2) and being dissipated there, $I_{1(2)}$ is the moment of inertia of WD 1(2), and $\dot{E}_{\rm pp}$ is the point-particle GW power.  

\begin{figure}[tb]
  \centering
  \includegraphics[width=\columnwidth]{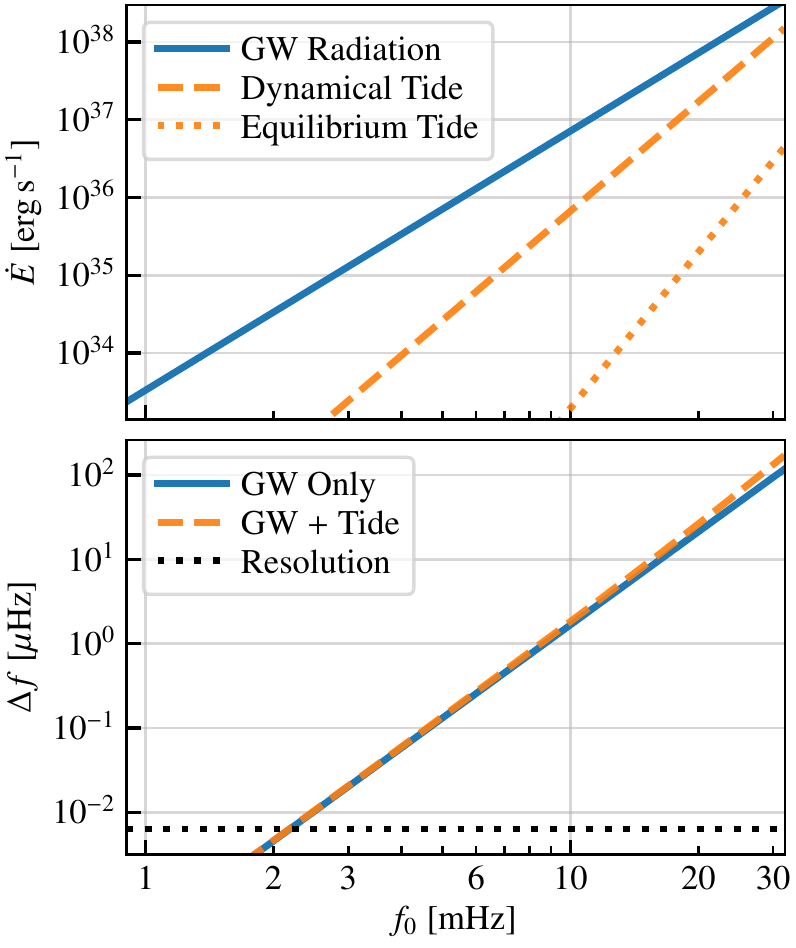}
\caption{Tidal interactions for a $0.72\,M_\odot$-$0.6\,M_\odot$ WD binary. Upper panel: orbital energy dissipation/transfer rates $\dot{E}$ in different channels. Lower panel: total GW frequency shift of the binary over a 5 year observation period, $\Delta f$, as a function of the initial GW frequency $f_0$. Frequency shifts greater than $1/T_\mathrm{obs}$ are resolvable.}
\label{fig:WDtide}
\end{figure}

In the top panel of Figure~\ref{fig:WDtide}, we show the energy dissipation rate via different channels as a function of the system's GW frequency. Here we focus on a $(M_1,\ M_2)=(0.72\,M_\odot,\ 0.6\,M_\odot)$ WD binary. We compute the radii using Eq.~(\ref{eq:RvsM}) and assume $I_{1(2)}=0.26M_{1(2)}R_{1(2)}^2$. When the system enters TianGO's most sensitive band of $f>10\,{\rm mHz}$, the dynamical tide accounts for more than $10\%$ of the orbital energy loss. As a comparison, the energy transferred into the equilibrium tide (as computed following Ref.~\cite{Burkart2013}) is only a minor amount. 

The tidal interaction accelerates the orbital decay and thus increases the amount of frequency chirping during a given period, as is illustrated in the bottom panel of \cref{fig:WDtide}. In the plot we show the increase in system's GW frequency over an observation period of 5 years with (the orange trace) and without (the blue trace) the tidal effect as a function of the initial frequency $f_0$ at the start of the observation. Note that $\dot{E}_{\rm tide1(2)}\propto I_{1(2)}$. Therefore, measuring the excess frequency shift will allow us to directly constrain the moment of inertia of WDs. 

To quantify the detectability of $I_{1(2)}$, we construct GW waveforms taking into account the tidal interactions (see \cref{sec:TidalWaveforms} for details) and then use the Fisher matrix to estimate the parameter estimation error. We focus on the same $(M_1,\ M_2)=(0.72\,M_\odot,\ 0.6\,M_\odot)$ WD binary as before and fix its distance to be 10\,kpc but randomize its orientation and sky location. The median uncertainty in WD's moment of inertia over a 5-year observation is summarized in Table~\ref{tab:WDdeltaI} for different initial GW frequencies. Due to the way the moment of inertia enters the waveform, we are most sensitive to the sum $(I_1 + I_2)$ and it can be constrained to a level of better than $1\%$ for sources at a gravitational-wave frequency of $f > 10\,{\rm mHz}$. 

\begin{table}
 \caption{Uncertainties in the sum of WDs' moment of inertia for different initial GW frequencies at the start of a 5 year observation period.}
 \label{tab:WDdeltaI}
 \begin{ruledtabular}
 \begin{tabular}{ccccc}
 $f_0$ [mHz]                                      &  5  & 10 & 20 & 30  \\
 \hline
$\frac{\Delta (I_1 + I_2)}{(I_1+I_2)}$       &   1.1 & $3.3\times10^{-3}$ & $9.6\times10^{-6}$ & $6.7\times10^{-7}$
 \end{tabular}
 \end{ruledtabular}
 \end{table}

With such a high level of statistical accuracy, we can imagine that a precise relation between WD's mass and moment of inertia can be established after a few detections. We can then use this tidal effect to improve the measurability of other parameters. For example, due to a WD binary's  slow orbital motion --- $\left(v_{\rm orb}/c \right)^2< 10^{-4}$ even at the onset of Roche-lobe overflow, where $v_{\rm orb}$ is orbital velocity --- it is challenging to measure parameters such as the mass ratio that come from high-order post-Newtonian corrections using the point-particle GW waveform alone. However, it is critical to know not only the chirp mass but also the component masses when tackling problems like identifying progenitors of type Ia supernovae (Section~\ref{sec:typeIa}). Nonetheless, if we assume $I{=}I(M_{\rm wd})$, the tide will then introduce a mass dependence that is different from the chirp mass and has a more prominent effect on the orbital evolution than the post-Newtonian terms. It is thus a promising way to help constrain a WD binary's component masses. 

\begin{figure}[tb]
  \centering
  \includegraphics[width=\columnwidth]{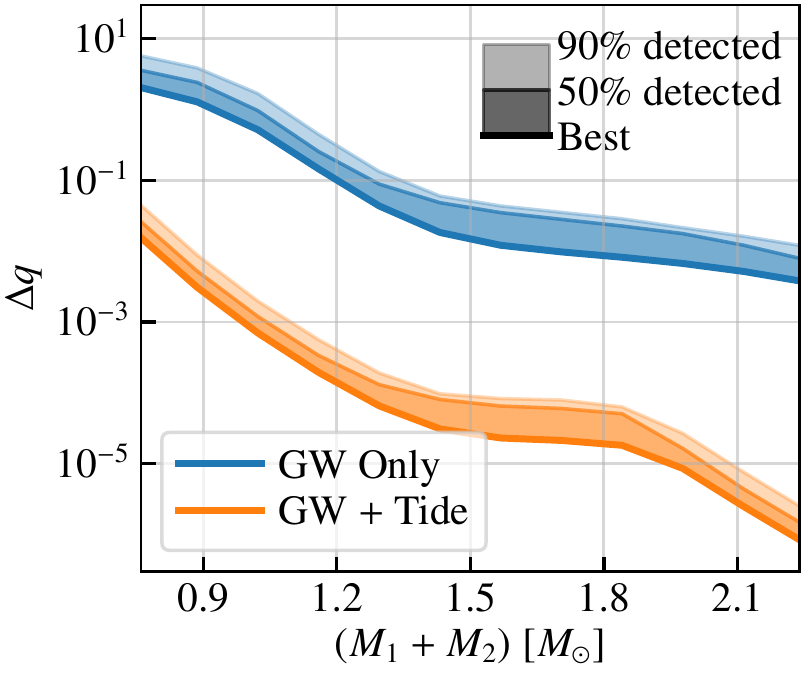}
\caption{Uncertainties in inferring the mass ratio, $\Delta q$,  for WD binaries with different total masses. }
\label{fig:PEwTide}
\end{figure}

This is illustrated in Figure~\ref{fig:PEwTide}. Here we compare the parameter estimation uncertainty on the mass ratio for systems with different total masses. We set each system's GW frequency to be the one right before the Roche-lobe overflow and fix the true mass ratio to be 1.2. 
When tides are included, we assume a fixed relation between a WD's moment of inertia and mass as 
\begin{equation}
I(M_{\rm wd}) = 3.1 \times 10^{50}  \left(\frac{M_{\rm wd}}{0.6\,M_\odot}\right)^{1/3}\, {\rm g\,cm^{2}}.
\end{equation}
Compared to the point-particle results (blue traces), the ones including the tidal effect (orange traces) can reduce the statistical error on mass ratio, $\Delta q$, by nearly three orders of magnitude over a large portion of parameter space. 

\section{Constraining Progenitors of  Black Hole Binaries by measuring spins}
\label{sec:BHspin}

The detections by aLIGO and aVirgo have confirmed the existence of stellar-mass BH binaries. A question to ask next is then what is the astrophysical process that gives birth to these systems. Currently, the two most compelling channels are isolated binary evolution in galactic fields~\cite{Belczynski2016, Mandel2016} and dynamical formation in dense star clusters~\cite{Rodriguez2015}. A potentially powerful discriminator of a system's progenitor is the spin orientation (see, e.g., Refs.~\cite{Rodriguez2016, Farr2017, Stevenson2017, Farr2018, Liu2018}). Isolated field binaries will preferentially have the spin aligned with the orbital angular momentum, whereas in the case of dynamical formation the orientation is more likely to be isotropic. 

While ground-based detectors are sensitive to the {\it effective aligned spin parameter} $\chieff$ (the mass-weighted sum of two BHs' dimensionless spins along the direction of orbital angular momentum~\cite{Hannam2014}), the determination of spin components that lie in the orbital plane, often parameterized as the {\it effective precession spin parameter} \chip~\cite{Schmidt2015}, will be challenging due to the limited sub-10\,Hz sensitivities for ground-based detectors~\cite{Yu2018}. TianGO, on the other hand, is sensitive down to $10\,{\rm mHz}$ and can thus measure the modulations due to the precession spin $\chip$ with much higher accuracy. TianGO thus allows us to construct a \emph{two-dimensional} spin distribution (in $\chieff$ and $\chip$) of stellar-mass BH binaries that cannot be constructed with ground-based detectors alone, and consequently provide valuable insights into the formation history of binaries. 

\begin{figure}[tb]
  \centering
  \includegraphics[width=\columnwidth]{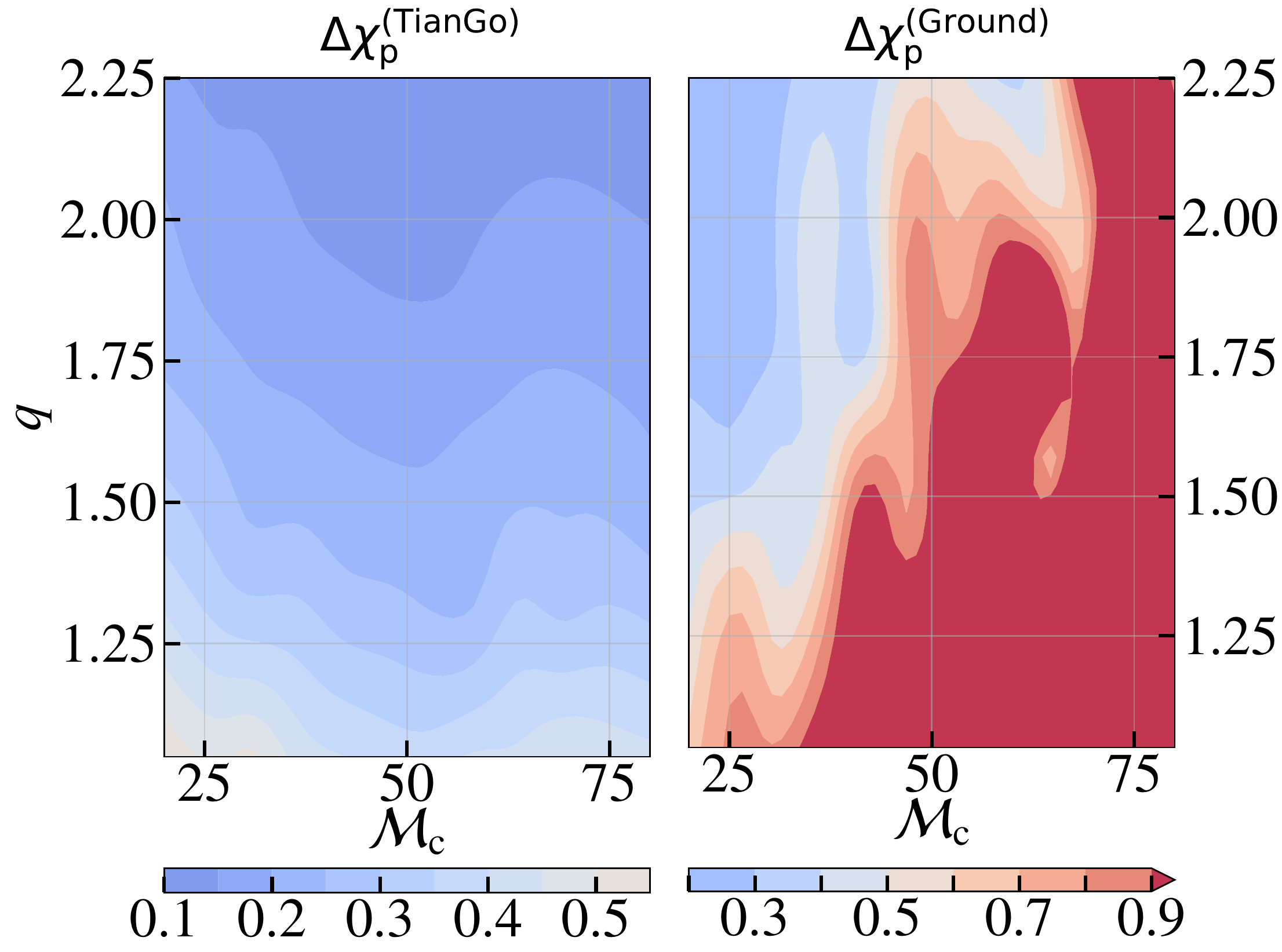}
  \caption{Uncertainties in the precession spin parameter $\chip$ for TianGO (left) and a network of five Voyager-like detectors (right). We vary the source's chirp mass and mass ratio, while fixing $(\chieff,\,\chip)$=$(-0.3,\,0.6)$. The source is assumed to be at $z=2$ and the sky location is marginalized over. Note that the color scales are different in the two panels.
  }
\label{fig:chi_p}
\end{figure}

In Figure~\ref{fig:chi_p}, we show the sky-location-averaged uncertainty in $\chip$ for sources located at a redshift of $z=2$ ($D_{L}\simeq 16\,{\rm Gpc}$). To capture the precession effect, we use the \texttt{IMRPhenomPv2} waveform model~\cite{Hannam2014} and assume all sources to have a moderate spin rate of $(\chieff,\,\chip)$=$(-0.3,\,0.6).~$\footnote{Specifically, here we set the components of the spins as $\chi_{1z}=\chi_{2z}=\chieff$, $\chi_{1x}=\chi_{2x}=\chip$, and $\chi_{1y}=\chi_{2y}=0$\label{fn:spin}. The (non-unique) way of choosing the components does not significantly affect the final results, as these components only enter the inspiral part of the waveform via the combination $(\chieff,\,\chip)$ in the \texttt{IMRPhenomPv2} waveform.  The initial frequency we choose to set the spin components is fixed at 0.01\,Hz, consequently fixing the orbital and spin precession phases.} These values are chosen for illustrative purposes, yet the conclusions we draw are generic. The source-frame chirp mass $\mathcal{M}_c$ and mass ratio $q$ are allowed to vary.   As shown in the figure, for TianGO (left panel), $\chip$ is measurable ($\Delta \chip < |\chip|$) in almost the entire parameter space as long as the mass ratio is slightly greater than 1. As a comparison, a network of ground-based detectors consisting of HLVKA (right panel), can only detect $\chip$ over a small portion of the parameter space ($\mathcal{M}_c<40\,M_\odot$ and $q>1.4$). This demonstrates TianGO's unparalleled ability to determine $\chip$.  

One caveat though is that the above analysis assumes binary BHs have a broad range of spins with $0.1\lesssim a/M <1$ as in the case of X-ray binaries~\cite{Miller2015}. However, the BBHs detected by aLIGO and aVirgo during the first and second observing runs~\cite{lsc2018a} suggest that most BHs may have only low spins of $a/M<0.1$\footnote{Ref.~\cite{Zackay2019} reported a highly spinning BBH, yet this event has lower detection significance compared to the others. If the event is indeed astrophysical, it might hint at a chemically homogeneous formation~\cite{Mandel2016}.}~\cite{lsc2018b}, which may be the consequence of an efficient angular momentum transfer in the progenitor stars~\cite{Fuller2019b}. In this case, a moderate $\chip$ would be an indication of the merger event involving a second-generation BH~\cite{Rodriguez2018}. 

\begin{figure}[tb]
  \centering
  \includegraphics[width=\columnwidth]{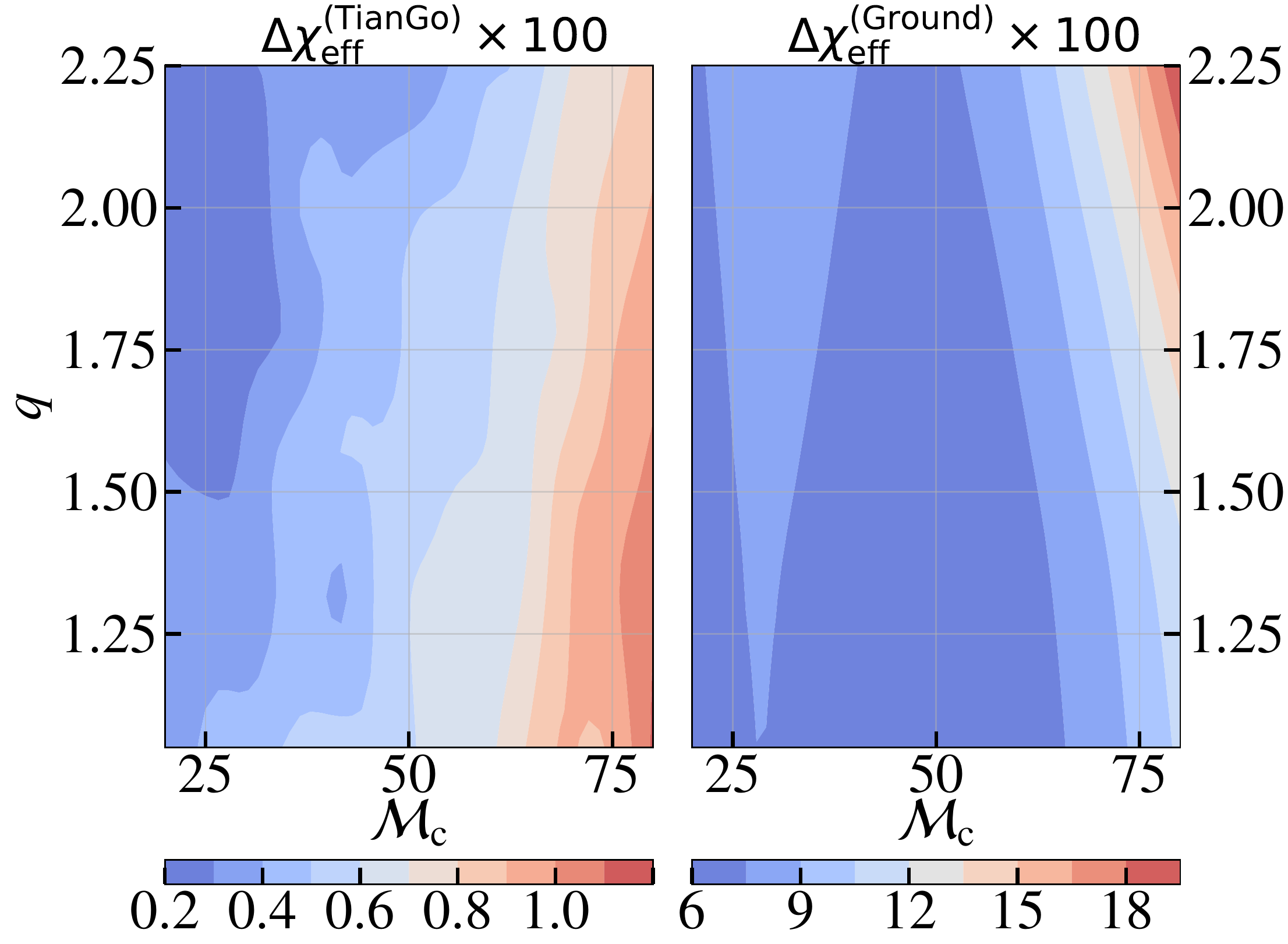}
\caption{Similar to Figure~\ref{fig:chi_p} but showing the uncertainties in \chieff, for binaries with  $(\chieff,\,\chip)$=$(0.05,\,0)$.  Note that the uncertainties $\Delta \chieff$ are amplified by a factor of 100 in the plots, and that the color scales are different in the two panels.}
\label{fig:chi_eff}
\end{figure}

As for the majority of the slowly spinning BHs, TianGO can still deliver valuable information ground-based detectors cannot access. This is illustrated in Figure~\ref{fig:chi_eff} where we present the uncertainty in $\chieff$. This time we assume the system to only have a slow spin rate of $(\chieff,\,\chip)$=$(0.05,\,0)$ while the other parameters are the same as in Figure~\ref{fig:chi_p}. The Voyager network cannot constrain $\chieff$ for systems spinning at such a slow rate. TianGO, on the other hand, can still achieve an accuracy of $\Delta \chieff / \chieff \lesssim 0.3$ over most of the parameter space. This opens up the possibility of discriminating different angular momentum transfer models that all predict the majority of BHs having spins in the range of $a/M\sim 0.01-0.1$~\cite{Spruit2002, Heger2005, Qin2018, Fuller2019a, Fuller2019b, Bavera2019}. 

\section{Revealing orbital eccentricity evolution}
\label{sec:eccOrb}
So far our discussions have focused on systems with circular orbits. This is a good assumption for signals at $f>10\,{\rm Hz}$ as the GW radiation may have efficiently dissipated away the initial eccentricity. Nonetheless, at lower frequencies the residual eccentricity left from the binary's formation may leave a detectable imprint on the GW waveform. While LISA can detect a fraction of the eccentric systems at a few tens of millihertz if the initial eccentricity is mild (see, e.g., Refs.~\cite{Breivik2016, Nishizawa2016}), it will likely miss those formed with very high initial eccentricities of $(1{-}e_0){\lesssim} 0.01 $~\cite{Chen2017}. Such a high initial eccentricity can be produced if the binary is formed via binary-single scattering~\cite{Samsing2014, Amaro-Seoane2016}, hierarchical triple interactions~\cite{Antonini2014, Naoz2016, Liu2018, Liu2019}, or gravitational braking~\cite{OLeary2009}; see Ref.~\cite{Chen2017} for a comprehensive summary. A decihertz detector like TianGO will then be the only way to detect the evolution of such systems. We elaborate on this point further in this section.


The detectability of the orbital eccentricity has been studied in detail in Ref.~\cite{Barack2004} whose key components are summarized in the following. The GW strain from an eccentric binary can be decomposed into a superposition of different orbital harmonics as 
\begin{equation}
h(t) = \sum_{k=1}^{\infty} h_k (t),
\end{equation}
where each harmonic varies at a frequency $f_k$ given by 
\begin{equation}
f_k = k \Omega_{\rm orb}/2\pi  + \dot{\gamma}/\pi.
\label{eq:f_k}
\end{equation}
The angle $\gamma$ represents the direction of the pericenter, and we have defined $f_k$ as the average between the radial and azimuthal frequencies. Note that for a circular orbit, all the GW power is radiated through the $k=2$ harmonic in the leading-order quadrupole approximation, and hence there exists a one-to-one mapping between time and GW frequency [cf.~\cref{eq:t_merger}]. When the orbit is eccentric, however, at a given instant the GW strain contains multiple frequency components. 

In the frequency domain, the characteristic strain amplitude $h_{c,k}(f_k)$ of harmonic $k$ is~\cite{Barack2004}\footnote{Note that $h_{c,k} (f_k)$ is a dimensionless quantity.}
\begin{equation}
h_{c, k} (f_k) = \frac{1}{\pi D_{L}} \sqrt{\frac{2G}{c^3}\frac{\dot{E}_k}{\dot{f}_k}},
\label{eq:h_ck}
\end{equation}
where $\dot{E}_k$ is the GW power radiated at frequency $f_k$ (see Ref.~\cite{Peters1963}). The corresponding SNR for each harmonic can then be evaluated as 
\begin{equation}
\left\langle {\rm SNR}_k^2 \right\rangle = \int \frac{h_{c,k}^2(f)}{5  f S_a(f)} \diff \ln f,
\end{equation}
where $S_a(f)$ is the power spectral density of the noise in detector $a$, and the factor of $5$ in the denominator accounts for the averaging over sky-location. To evaluate \cref{eq:h_ck} we first integrate the Keplerian elements $(\Omega_{\rm orb},\,e,\,\gamma)$ (i.e., the orbital frequency, eccentricity, and argument of pericenter, respectively)  from a set of initial values to the final plunge. We then evaluate $\dot{E_k}\left[ f_k (t)\right]$ and $\dot{f}_k\left[ f_k (t)\right]$ at each instant $t$ with the corresponding Keplerian elements at the same moment under a post-Newtonian approximation as was done in Ref.~\cite{Barack2004}. The spin has been neglected throughout this section. 

\begin{figure}[tb]
  \centering
  \includegraphics[width=\columnwidth]{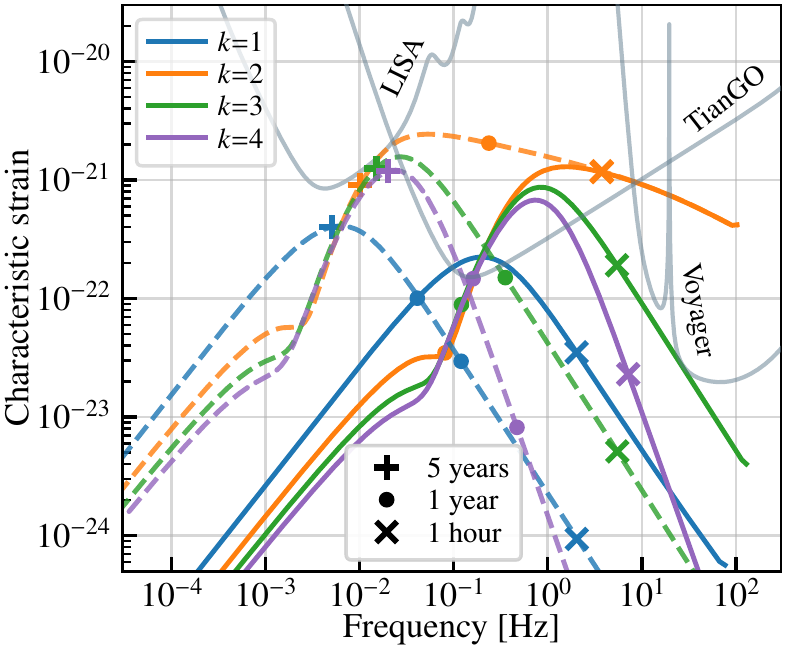}
\caption{Evolution of the characteristic strain amplitude $h_c$ for the first four orbital harmonics of a BH binary with $M_1{=}M_2{=}30\,M_\odot$ at $z{=}0.3$. The solid (dashed) curves represent systems with an initial eccentricity of $1{-}e_0{=}10^{-3} (10^{-2})$ and $a_0{=}0.1\,{\rm AU}$. The x-axis corresponds to the frequency of each harmonic [\cref{eq:f_k}].  
The thin black traces represent the sky-averaged sensitivity (in $\sqrt{5fS_a(f)}$) for LISA, TianGO, and (a single) Voyager, respectively. The pluses, dots, and crosses correspond, respectively, to instants 5 years, 1 day, and 1 hour prior to the final merger.}
\label{fig:ecc_SM}
\end{figure}

%
%
In \cref{fig:ecc_SM} we show the evolution of the first four harmonics for a system with $M_1{=}M_2{=}30\,M_\odot$ at a redshift of $z{=}0.3$. We consider two initial conditions. The solid lines correspond to a system formed with an initial semi-major axis and eccentricity of $(a_0,\, 1{-}e_0){=}(0.1\,{\rm AU},\,10^{-3})$\footnote{This is the same system as the highly eccentric binary considered in figure 1 of Ref.~\cite{Chen2017} except for that we place the system at a further distance.}, and the dashed lines correspond to $(a_0,\, 1{-}e_0){=}(0.1\,{\rm AU},\,10^{-2})$. Such systems can form via triple interactions in dense stellar environments. On each curve we also mark the times corresponding to (5 years, 1 day, 1 hour) prior to the final plunge with the (plus, dot, cross) symbols. As shown in the plot, systems with such high initial eccentricities are likely to be missed by LISA because when the system has an orbital frequency of $\Omega_{\rm orb}/2\pi {=} \text{a few}\times {\rm mHz}$, a significant amount of the total GW power is radiated through the third and forth (and even higher) harmonics whose GW frequencies $f_k {\simeq} k\Omega_{\rm orb}/2\pi$ are higher than LISA's most sensitive band. Meanwhile, the amplitude of the $k {\geq} 3$ harmonics decays quickly as the eccentricity is damped by the GW radiation, and become negligible in the ${>}10\,{\rm Hz}$ band. Nevertheless, the $k{=}2-4$ harmonics all have significant amplitudes in the 0.01-1 Hz band, making TianGO the ideal instrument to study the orbital evolution and further constrain the binary's formation channel (in addition to spin which we studied in \cref{sec:BHspin}).  

\begin{figure}[tb]
  \centering
  \includegraphics[width=\columnwidth]{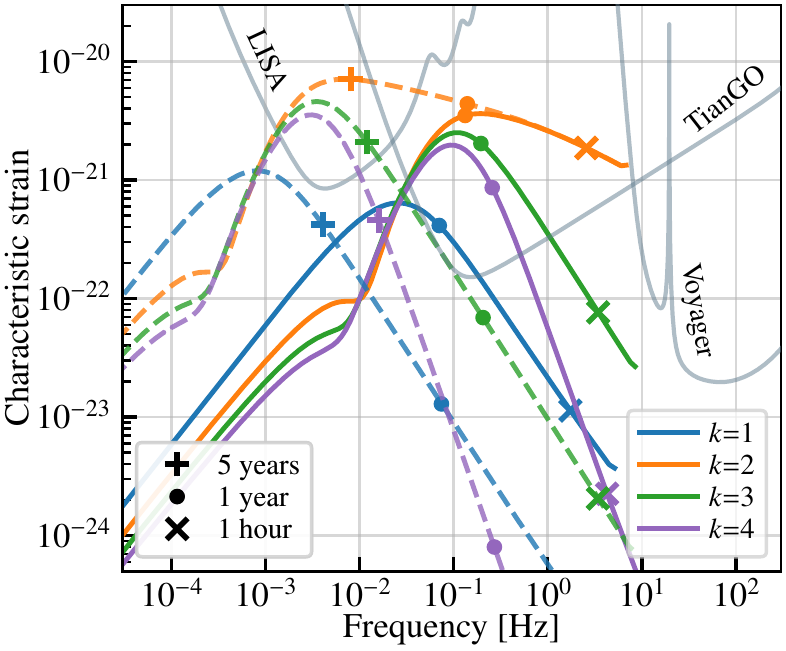}
\caption{Similar to \cref{fig:ecc_SM}, but for a IMRI system with $(M_1,\,M_2){=}(10^3\,M_\odot,\,10\,M_\odot)$ starting at $a_0{=}1\,{\rm AU}$. The solid (dashed) traces correspond to an initial eccentricity of $1-e_0=10^{-3} (10^{-2})$. }
\label{fig:ecc_IMBH}
\end{figure}

In addition to the dynamical formation of stellar-mass BH binaries, an IMBH capturing a stellar-mass BH (i.e., the IMRI systems we discussed in \cref{sec:IMBH}) can also form with high initial eccentricity of $(1{-}e_0){\sim}10^{-3}$~\cite{Konstantinidis2013, Leigh2014, Amaro-Seoane2018}. The detectability of such systems are studied in Figure~\ref{fig:ecc_IMBH}. Here we focus on a binary with $(M_1,\,M_2){=}(10^3\,M_\odot,\,10\,M_\odot)$. We also change the initial conditions to be $(a_0,\, 1{-}e_0){=}(1\,{\rm AU},\,10^{-3})$ for the solid traces and $(a_0,\, 1{-}e_0){=}(1\,{\rm AU},\,10^{-2})$ for the dashed ones to be consistent with Ref.~\cite{Amaro-Seoane2018}. The other parameters, including the meanings of symbols, are the same as those used in Figure~\ref{fig:ecc_SM}. Broadly speaking, LISA favors detecting systems formed at large initial separation with low eccentricity while TianGO is more suitable to observe those formed at small separation with high eccentricity. Once again, we see that TianGO would be necessary to cover the entire parameter space of $(a_0, 1{-}e_0)$.

\section{Inferring the presence of tertiary masses near merging binaries}
\label{sec:third_mass}
Highly eccentric orbits, such as those discussed in the previous section, are likely a consequence of the presence of a tertiary mass near the coalescing binary. In this section, we discuss a direct method to infer the existence of this third body. In such a three-body system, TianGO detects the GW radiation from the inner binary. It is thus possible to infer the presence of the third body since 
the Keplerian motion of the outer orbit (formed by the tertiary mass and the center of mass of the inner binary)
causes a time-varying Doppler shift to the inner binary's waveform~\cite{Mandel2018}. 
Studies have shown (see, e.g., Ref.~\cite{Bartos2017}) that a dense population of stellar-mass BH binaries may reside in galactic nuclei. Accretion discs near the central supermassive BHs can further accelerate the mergers of stellar-mass BH binaries by providing gaseous torques, hence making those binaries a significant fraction of the population detected by GW observatories. Meanwhile, the central supermassive BHs also serve as the tertiary bodies whose gravitational fields modulate the inner binary's waveform. We thus study the detectability of such modulations. 

In this section, we focus on the last stage of the inner binary's evolution where the GW radiation dominates the energy loss, and thus ignore the gaseous effects. In addition to the GW-driven evolution, the inner binary's phase in the barycentric frame is further modulated by its orbit around the supermassive BH (which we refer to as the outer orbit). In the simplest case where the inner and outer orbits are coplanar and the outer orbit is circular, the GW phase is given by~\cite{Randall2019} 
\begin{equation}
\phi_{\rm bc} (t) = 2\pi \int f(t) \left[1 - \frac{v_{\rm mod}}{c}\sin(\Omega_{\rm mod}t + \phi_{\rm mod} )\right] \diff t,
\label{eq:phi_bc}
\end{equation}
where $v_{\rm mod}$, $\Omega_{\rm mod}$, and $\phi_{\rm mod}$ are, respectively, the orbital velocity, angular velocity, and initial phase of the outer orbit. The subscript ``bc'' indicates that the GW phase $\phi_{\rm bc}$ is in the barycentric frame. From \cref{eq:phi_bc}, it is easy to see that the orbital acceleration around the tertiary mass acts as a frequency modulation of the GW waveform. 
Numerically,
\begin{align}
&\frac{v_{\rm mod}}{c} = 0.02  \left(\frac{M_3}{4\times10^6\,M_{\odot}}\right)^{1/2} \left(\frac{r_3}{100\, {\rm AU}}\right)^{-1/2},\\
&\frac{\Omega_{\rm mod}}{2\pi} = \frac{1}{0.5\,{\rm yr}} \left(\frac{M_3}{4\times10^6\,M_{\odot}}\right)^{1/2} \left(\frac{r_3}{100\,{\rm AU}}\right)^{-3/2},
\end{align}
where $M_3$ is the mass of the tertiary body and $r_3$ is the distance between it and the center of mass of the inner binary.

To estimate the detectability of this modulation, we follow the approach of Ref.~\cite{Randall2019} and perform a time-domain Fisher matrix analysis (see Ref.~\cite{Barack2004} for details). 
As in Ref.~\cite{Randall2019}, we construct the intrinsic waveforms using only the leading order quadrupole formula parameterized by the chirp mass of the inner binary $\mathcal{M}_c$, the initial time $t_0$, and the initial phase  $\phi_0$. We further add phase modulations to the intrinsic waveform according to Eq.~(\ref{eq:phi_bc}). Since we assume that the outer orbit is circular and coplanar with the inner orbit, this introduces the additional three parameters $v_{\rm mod}$, $\Omega_{\rm mod}$, and $\phi_{\rm mod}$. The time-dependent effects of the TianGO or LISA orbits discussed in \cref{sec:PE-overview}, and the effects of higher order post-Newtonian corrections to the waveform are deffered to future studies. The sky-averaged sensitivities are used to normalize the strain.

\begin{figure}[tb]
  \centering
  \includegraphics[width=\columnwidth]{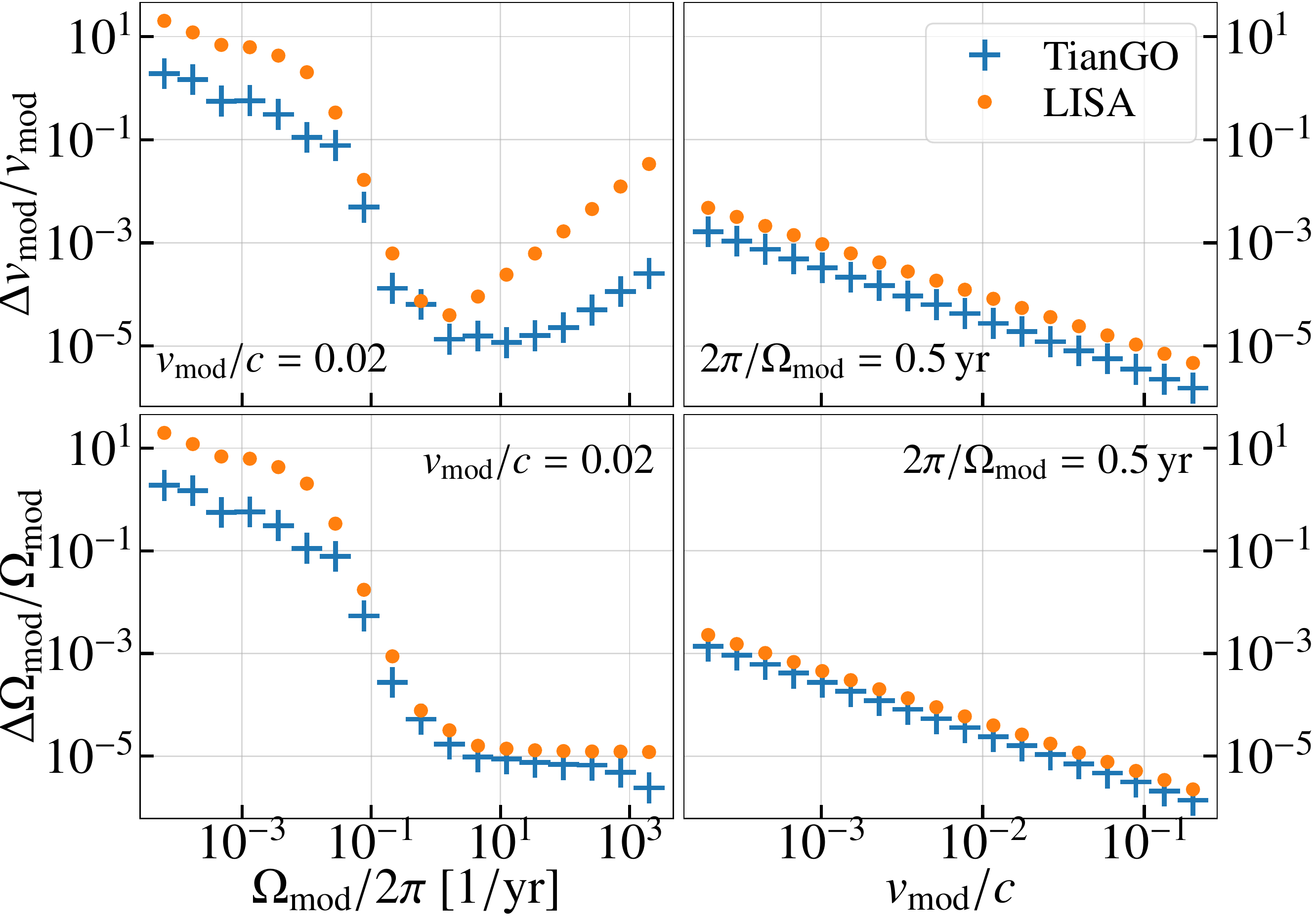}
  \caption{Fractional uncertainties in the linear velocity (top panels) and angular velocity (bottom panels) of the outer orbit for an inner binary with masses $(30\,M_{\odot}, 30\,M_{\odot})$ and an initial frequency of $0.017\,{\rm Hz}$, corresponding to a system that merges in 5 year. The system is $400\,\mathrm{Mpc}$ from the earth and $\phi_{\rm mod} = 0$. In the left panels we fix $v_{\rm mod}/c=0.02$ and vary $\Omega_{\rm mod}$, and in the right panels we fix $\Omega_{\rm mod} = 2\pi / 0.5 \,{\rm yr}$ and vary $v_{\rm mod}$. Blue pluses and orange dots represent the uncertainties obtained with TianGO and LISA, respectively.
  }
\label{fig:pe_third_obj}
\end{figure}

The existence of the tertiary mass could be detected if the fractional errors $\Delta v_{\rm mod}/v_{\rm mod}$ and $\Delta \Omega_{\rm orb}/\Omega_{\rm orb}$ are both less than unity. We demonstrate the detectability of this phase modulation for a system $400\,\mathrm{Mpc}$ from the Earth in \cref{fig:pe_third_obj}. Here we focus on an inner binary with masses $(30\,M_{\odot}, 30\,M_{\odot})$ and an initial frequency of $0.017\,{\rm Hz}$. This initial frequency is chosen so that the binary merges over a 5-year mission. In the left panels we fix the outer orbit's linear velocity to be $v_{\rm mod}/c{=}0.02$ and vary its angular velocity $\Omega_{\rm mod}$, while in the right panels $\Omega_{\rm mod}$ is fixed to $2\pi / 0.5 \,{\rm yr}$ and $v_{\rm mod}$ is varied. If the third mass is a supermassive BH with $M_3{=}4\times10^6\,M_\odot$, the range of $\Omega_\mathrm{mod}$ and $v_\mathrm{mod}$ shown in the figure corresponds to changing the inner binary's distance to $M_3$ from 1\,AU to $10^5\,{\rm AU}$.

As shown in \cref{fig:pe_third_obj}, TianGO (blue pluses) has a peak sensitivity to the outer orbit's modulation 2-3 times better than that of LISA (orange dots). The fractional uncertainties in both $v_{\rm mod}$ and $\Omega_{\rm mod}$ can be constrained to the level of $10^{-5}$ if the tertiary mass is a Sgr A$^\ast$-like supermassive BH with $M_3 \sim 4\times10^6\,M_\odot$ and distance to the inner binary of $r_3 \sim 100\,{\rm AU}$. Meanwhile, as $\Omega_{\rm orb}$ moves away from $\sim 2\pi/1\,{\rm yr}$, TianGO can outperform LISA even more. Indeed, since TianGO has an overall better sensitivity to stellar-mass BH binaries (i.e., the inner binaries) than LISA (cf.\ Fig.~\ref{fig:horizons}), it also has better sensitivity to excess modulations on the GW waveforms emitted by those binaries. 

Note that while the discussion above focused on the case where the tertiary mass is a supermassive BH ${\sim}100\,{\rm AU}$ from the inner binary, the results are the same if it is a stellar-mass object a few AU away with similar $v_{\rm mod}$ and $\Omega_{\rm mod}$. Furthermore, this method can be applied to the search for exoplanets around WD binaries, as suggested by  Refs.~\cite{Seto2008, Tamanini2019}, since an exoplanet in such a system acts as a tertiary mass modulating the WD binary's phase.

\section{Conclusion and Discussions}
\label{sec:conclusion}
The scientific rewards of having a combined network of ground- and space-based GW detectors is greater than the sum of its parts. In particular, the very long baseline achieved by coherently combining detectors will allow for an unprecedented ability to estimate the parameters of compact binaries.

We showed that TianGO can significantly enhance the sky localization and distance estimation accuracy of binary BHs when combined with a network of ground-based detectors. This has the potential to resolve the tension between the local and cosmological measurements of the Hubble constant. It can also localize a coalescing NS binary with unparalleled accuracy weeks before the final merger, consequently facilitating multi-messenger astronomy. Exploring the evoloution of binary white dwarfs, in combination with LISA and TianQin, may yield critical insights into the type-1a SNe progenitor mystery.

Meanwhile, this mission will also distinguish the formation channels of stellar-mass BH binaries by measuring both the spins and the orbital eccentricities. If such a binary merges under the influence of a tertiary mass, TianGO may directly probe the existence of the perturber by measuring the Doppler phase shifts in the GW waveform. It also helps constrain the formation history of today's massive BHs and the search for IMBHs. 

In addition to the astrophysics described in the main text, there are many other discoveries TianGO may enable. For example, TianGO may also enable a direct detection of the GW memory effect~\cite{Christodoulou:91, Favata:09} or test the angular distribution of the memory background-based~\cite{Yang2018}. If TianGO does detect IMBHs, we may further use them to search for or rule out the existence of ultra-light bosons via superradiance~\cite{East:17, Brito2017}. We plan to carry out more in depth studies of TianGO's scientific capabilities in the future.

\section*{Acknowledgments}
\label{sec:Ack}
We would like to thank Baoyi Chen, Curt Cutler, Michael Coughlin, Tom Callister, Carl Haster, Jameson Rollins, Evan Hall, Jim Fuller, Rory Smith, Salvatore Vitale, Will Farr, and Jan Harms for discussions.  KAK and RXA were supported by Boeing (Award Number CT-BA-GTA-1). HY is supported by the Sherman Fairchild Foundation. YC is supported by NSF grant PHY-1708213 and by the Simons Foundation (Award Number 568762).

\appendix
\section{Parameter Estimation with a Combined Network of Space and Ground-Based Detectors}
\label{sec:PE-overview}

The calculations done in this paper use the well-known Fisher matrix formalism~\cite{Finn1992, Cutler1994, Vallisneri2008} which we briefly summarize here. We then explain the methods used to simultaneously analyze combined network of both space and ground-based gravitational wave detectors.

Suppose the frequency domain signal measured in the detector $a$ is
\begin{equation}
s_a(f) = h_a(f, \vect{\theta}) + n_a(f)
\end{equation}
where $h_a(f, \vect{\theta})$ is the gravitational wave waveform and $n_a$ is stationary Gaussian noise with single-sided power spectral density $S_a(f)$. The waveform depends on a set of parameters $\vect{\theta}$ that are to be inferred from the measurement of $s_a$. For large signal to noise ratios, the differences $\Delta \theta_i$ between the measured and true parameters, as measured by detector $a$, are normally distributed
\begin{equation}
p(\Delta\theta) \propto \mr{e}^{-\Gamma^a_{ij}\Delta\theta_i \Delta\theta_j/2}
\end{equation}
where
\begin{equation}
\Gamma^a_{ij} = \ip{\frac{\pd h}{\pd \theta_i}}{\frac{\pd h}{\pd \theta_j}}_a
\end{equation}
is the Fisher information matrix and
\begin{equation}
\ip{g}{h}_a = 4\re \int_0^\infty \frac{g^*(f)h(f)}{S_a(f)}\,\mr{d} f
\end{equation}
is the noise weighted inner product for detector $a$.

The covariance for estimating the parameters $\theta$ with a network of detectors is the inverse of the network Fisher matrix obtained by summing the individual Fisher matrices
\begin{align}
\Gamma_{ij} &= \sum_a \Gamma_{ij}^a \nonumber \\
\Sigma_{ij} &= \left\langle \Delta\theta_i \Delta\theta_j\right\rangle
= \left(\Gamma^{-1}\right)_{ij}.
\end{align}
The angular uncertainty in determining a source's sky location is
\begin{equation}
\Delta\Omega = 2\pi\left|\cos\delta\right|
\sqrt{\Sigma_{\alpha\alpha} \Sigma_{\delta\delta}
- \left(\Sigma_{\alpha\delta}\right)^2},
\end{equation}
where $\alpha$ is the right ascension and $\delta$ is the declination.

\begin{figure}
  \centering
  \includegraphics[width=\columnwidth]{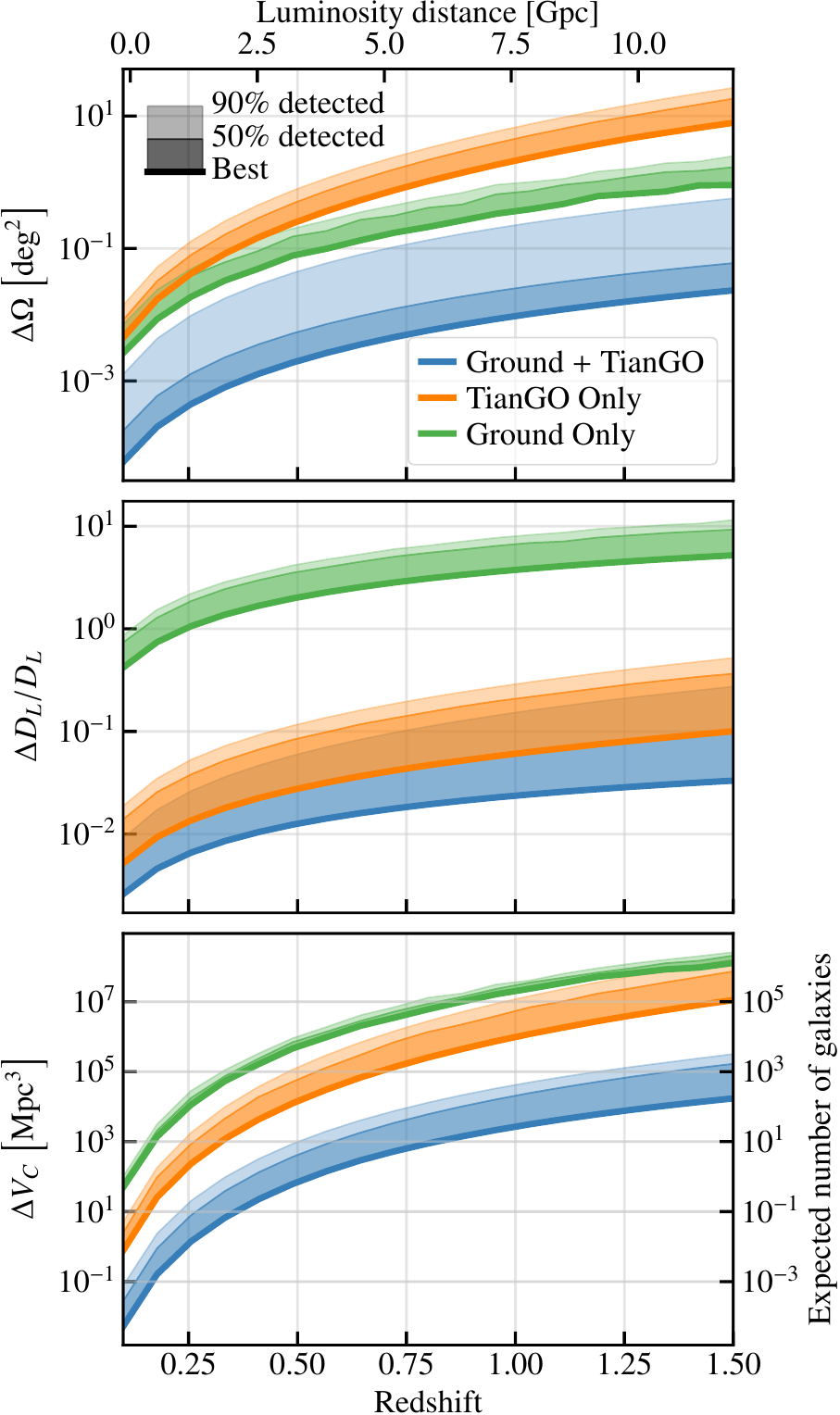}
\caption{The same as \cref{fig:sky-localization-30deg} except for face on binaries.}
\label{fig:sky-localization-0deg}
\end{figure}

To simultaneously describe the signal measured in both the ground-based and space-based detectors, especially since TianGO's sensitivity band extends well into that of the ground-based detectors (see \cref{fig:network}), one needs a waveform that captures both the high frequency merger-ringdown as well as the low frequency time-dependence associated with Doppler shifts and time-dependent antenna patterns. To do so, we modify the approach of Ref.~\cite{Cutler1998} to include phenomenological waveforms which include the merger and ringdown.

We first use \texttt{lalinference}~\cite{Veitch2015, lalsuite} to generate a phenomenological frequency domain waveform
\begin{equation}
u_\mr{ph}(f) = A_\mr{ph}(f)\,\mr{e}^{\mr{i}\Psi_\mr{ph}(f)}
\end{equation}
defined by the chirp mass $\mathcal{M}_c$, mass ratio $q$, luminosity distance $D_L$, and, where appropriate, the effective spin $\chi_\mr{eff}$ and spin procession $\chi_\mr{p}$. Except for in \cref{sec:BHspin} where the \texttt{IMRPhenomPv2} waveform~\cite{Hannam2014} is used, the \texttt{IMRPhenomD} waveform~\cite{Khan2016} is used throughout the paper. The source location is described by the right ascension $\alpha$ and declination $\delta$, and the orientation is described by the azimuthal and polar angles, $\phi_L$ and $\theta_L$, of the source's angular momentum $\rf{L}$. These remaining four extrinsic parameters as well as the coalescence time $t_c$ and phase $\phi_c$ are then added by hand as appropriate for the ground-based and space-based detectors. The coalescence time is defined as the time the wave arrives at the solar system barycenter.

For the ground-based detectors, the $+$ and $\times$ polarizations are first computed from $u_\mr{ph}$
\begin{subequations}
\begin{align}
h_+(f) &= A_\mr{ph}(f)\,\mr{e}^{\mr{i}\Psi_\mr{ph}(f)}\, \left(\frac{1 + \cos^2\iota}{2}\right) \\
h_\times(f) &= A_\mr{ph}(f)\,\mr{e}^{\mr{i}[\Psi_\mr{ph}(f) + \pi/2]}\,\cos\iota,
\end{align}
\end{subequations}
where $\iota$ is the source inclination. The signal observed in the ground-based detector $a$ is obtained by projecting $h_+$ and $h_\times$ onto the usual $+$ and $\times$ antenna patterns for that detector, $F^a_+(\alpha, \delta, \psi)$ and $F^a_\times(\alpha, \delta, \psi)$, where $\psi$ is the polarization phase.\footnote{
The conversion from $\phi_L$ and $\theta_L$ to $\psi$ and $\iota$ is
\begin{align*}
\cos\iota &= \cos\theta_L \sin\delta
+ \sin\theta_L \cos\delta \cos(\phi_L - \alpha)\\
\tan\psi &= \frac{\cos\theta_L + \cos\iota \sin\delta}
{\cos\delta\sin\theta_L\sin(\phi_L - \alpha)}.
\end{align*}
} See, for example, Appendix B of Ref.~\cite{Anderson2001}.

Finally, since the coalescence time is defined at the solar system barycenter, the phase is shifted by the light travel time $\tau_a(\alpha, \delta) = -\mathbf{d}_a \cdot \hat{\mathbf{n}}(\alpha,\delta)/c$ from the solar system barycenter to the detector where $\hat{\mathbf{n}}(\alpha,\delta)$ is the unit vector pointing from the barycenter to the source, and $\mathbf{d}_a$ is the vector from the barycenter to the detector. The waveform observed in detector $a$ is then
\begin{equation}
h_a(f) = \mr{e}^{\mr{i}\left[
2\pi f(t_c + \tau_a)
-\phi_c \right]}\left[F^a_+ h_+(f) + F^a_\times h_\times(f)\right].
\end{equation}

For TianGO, the time dependence of the antenna pattern and Doppler phase shift caused by the detector's orbit need to be included. The strategy employed by Ref.~\cite{Cutler1998}, which we follow, is to solve for the time dependence in the time domain and then, using a post-Newtonian expansion, find time as a function of frequency. This leads to the amplitude of the waveform being modulated by
\begin{equation}
\Lambda(f) = \sqrt{[1 + (\rf{L}\cdot\rf{n})^2]^2 F_+^2(f)
+ 4(\rf{L}\cdot\rf{n})^2 F_\times^2(f)}
\end{equation}
and gaining an extra phase
\begin{equation}
\tan \phi_p(f) = \frac{2(\rf{L}\cdot\rf{n}) F_\times(f)}
{[1 + (\rf{L}\cdot\rf{n})^2] F_+(f)}.
\end{equation}

Finally, as with the ground-based detectors, the phase associated with the propagation time from the solar system barycenter to TianGO must be included. Since sources can stay in TianGO's band for a significant portion of an orbit, this correction is the frequency dependent Doppler phase
\begin{equation}
\phi_D(f) = \frac{2\pi f}{c} R\cos\delta \cos\left[\phi_\mr{T}(f) - \alpha\right],
\label{Doppler-phase}
\end{equation}
where $\phi_\mr{T}(f)$ is the azimuthal angle of TianGO in its orbit around the sun and $R=1\,\mr{AU}$ is the radius of the orbit. As discussed in Ref.~\cite{Cutler1998}, higher order corrections to \cref{Doppler-phase} are of order $|\phi_D|(v/c)\lesssim 0.3\,(f/1\,\mr{Hz})$, where $v=2\pi R / T$ is the velocity of TianGO in its orbit and $T = 1\,\mr{yr}$. While this is not of concern for LISA, TianGO's sensitivity band extends past 10\,Hz where this correction becomes $\mathcal{O}(1)$. Future work can address the effects of this correction if necessary.

Putting it all together, the waveform observed by TianGO is
\begin{equation}
h_\mr{T}(f) = \Lambda(f) A_\mr{ph}(f)\,
\mr{e}^{\mr{i}\left[\Psi_\mr{ph}(f) - \phi_D(f) - \phi_p(f) - \phi_c \right]}.
\label{tiango-waveform}
\end{equation}
See \cref{sec:time-dependent-details} below for explicit expressions for $F_+(f)$, $F_\times(f)$, and $\phi_\mr{T}(f)$.

Fisher matrices are known to give unreliable results when waveforms that terminate abruptly in a detector's sensitivity band are used---as is often done when using inspiral only waveforms that terminate at the innermost stable circular orbit~\cite{Mandel2014}. Our analysis is not susceptible to such effects since we use a hybrid waveform that includes the merger and ringdown and which does not abruptly terminate.

\subsection{Explicit expressions for time-dependent waveforms}
\label{sec:time-dependent-details}

We collect here the important expressions from Ref.~\cite{Cutler1998} necessary to complete TianGO's waveform \cref{tiango-waveform}. The post-Newtonian expansion of Ref.~\cite{Cutler1998} is done in the parameter
\begin{equation}
x = \left(\frac{G}{c^3}\pi M_\odot\right)^{2/3} \frac{\mathcal{M}_c}{\mu}
\left[\left(\frac{\mathcal{M}_c}{M_\odot}\right) \left(\frac{f}{1\,\mr{Hz}}\right)\right]^{2/3}
\end{equation}
where $\mu = \mathcal{M}_c\,[q/(1+q)^2]^{2/5}$ is the reduced mass. The time as a function of frequency is
\begin{subequations}
\label{pn-time-freq}
\begin{equation}
t(f) \equiv t_f = t_c
- t_x \left[1 + \frac{4}{3}\left(\frac{743}{336}
+ \frac{11}{4}\frac{\mu}{M}\right) x - \frac{32\pi}{5}x^{3/2}\right],
\end{equation}
where
\begin{equation}
t_x = 5c^{5/3}\left(8\pi\, \frac{f}{1\,\mr{Hz}}\right)^{-8/3} \left(\frac{GM_\odot}{c^2}\right)^{-5/3}
\left(\frac{\mathcal{M}_c}{M_\odot}\right)^{-5/3},
\end{equation}
\end{subequations}
and $M$ is the total mass. The azimuthal angle of TianGO in its orbit around the sun is $\phi_\mr{T}(f) = 2\pi t_f/T$ where $T=1\,\mr{yr}$.

If $\phi=\alpha$, $\theta = \pi/2 - \delta$, and $\psi$ are the sky locations of a source and its polarization in the ecliptic frame, let $\tilde{\phi}$, $\tilde{\theta}$ and $\tilde{\psi}$ be those in the frame of the detector. Similarly, $\tilde{\phi}_L$ and $\tilde{\theta}_L$ are the azimuthal and polar angles of the source angular momentum $\rf{L}$ in the detector frame. If $\rf{z}$ is the unit vector along the $z$ direction, the polar angle of the source in the detector frame is
\begin{equation}
\cos \tilde{\theta}(t_f) = \rf{z}\cdot\rf{n}
= \frac{1}{2}\cos\theta - \frac{\sqrt{3}}{2} \sin\theta \cos(\phi_\mr{T}(f) - \phi),
\label{detframe-polar}
\end{equation}
the azimuthal angle of the source in the detector frame is
\begin{equation}
\tilde{\phi}(t_f) = \phi_\mr{T}(f)
+ \arctan\left[ \frac{\sqrt{3}\cos\theta + \sin\theta \cos(\phi_\mr{T}(t) - \phi)}
{2\sin\theta \sin(\phi_\mr{T}(f) - \phi)}\right],
\label{detframe-azimuthal}
\end{equation}
and the polarization phase of the source in the detector frame is
\begin{equation}
\tan\tilde{\psi}(t_f) = \frac{\rf{L}\cdot\rf{z} - (\rf{L}\cdot\rf{n})(\rf{z}\cdot\rf{n})}
{\rf{n}\cdot(\rf{L}\times\rf{z})}
\label{detframe-polarization}
\end{equation}
where
\begin{align}
\rf{L}\cdot\rf{z} &= \frac{1}{2}\cos\theta_L - \frac{\sqrt{3}}{2} \sin\theta_L \cos(\phi_\mr{T}(f) - \phi_L)\\
\rf{L}\cdot\rf{n} &= \cos\theta_L\cos\theta
+ \sin\theta_L \sin\theta \cos(\phi_L - \phi),
\end{align}
and
\begin{align}
\rf{n}\cdot(\rf{L}\times\rf{z}) &=
\frac{1}{2}\sin\theta_L \sin\theta \sin(\phi_L - \phi) \nonumber\\
&- \frac{\sqrt{3}}{2} \cos\phi_\mr{T}(f)\, \cos\theta_L \sin\theta \sin\phi \nonumber\\
&+ \frac{\sqrt{3}}{2} \cos\phi_\mr{T}(f)\, \cos\theta \sin\theta_L \sin\phi_L \nonumber\\
&- \frac{\sqrt{3}}{2} \sin\phi_\mr{T}(f)\,\cos\theta \sin\theta_L \sin\phi_L \nonumber \\
&+ \frac{\sqrt{3}}{2} \sin\phi_\mr{T}(f)\,\cos\theta_L \sin\theta \sin\phi.
\end{align}
\crefrange{detframe-polar}{detframe-polarization} are functions of frequency through \cref{pn-time-freq}.

The time-dependent antenna patterns as a function of frequency are given by plugging the detector frame angles \crefrange{detframe-polar}{detframe-polarization} into the standard antenna patterns
\begin{subequations}
\begin{align}
F_+(f) &= \left(\frac{1 + \cos^2\tilde{\theta}}{2}\right)\cos 2\tilde{\phi} \cos 2\tilde{\psi}
- \cos\tilde{\theta} \sin 2\tilde{\phi} \sin 2\tilde{\psi} \\
F_\times(f) &= \left(\frac{1 + \cos^2\tilde{\theta}}{2}\right)\cos 2\tilde{\phi} \sin 2\tilde{\psi}
+ \cos\tilde{\theta} \sin 2\tilde{\phi} \cos 2\tilde{\psi}.
\end{align}
\end{subequations}

\section{GW waveforms including tides in WDs}
\label{sec:TidalWaveforms}
In this section we derive the phase $\Psi(f)$ of the frequency-domain waveform $h(f)\propto \exp\left[ \imag \Psi(f)\right]$ including the effect of tidal interaction in WD binaries. 
As shown in Ref.~\cite{Cutler1994}, $\Psi(f)$ is related to the time-domain phase $\phi(t)$ as 
\begin{equation}
\Psi(f) = 2\pi f t(f) - \phi\left[t(f)\right] -\pi/4,
\end{equation}
we thus want to find how $t(f)$ and $\phi\left[t(f)\right]$ are modified by the tide. 

For the time as a function of frequency, we have
\begin{align}
t(f) & = \int \frac{\diff f}{\dot{f}} = \int\frac{\diff f}{\dot{f}_{\rm pp} + \dot{f}_{\rm tide}}, \nonumber\\
     &\simeq \int \frac{\diff f}{\dot{f}_{\rm pp}} \left(1 - \frac{\dot{E}_{\rm tide}}{\dot{E}_{\rm pp}}\right), \nonumber \\
     & = t_{\rm pp}(f) - \int \frac{1}{\dot{f}_{\rm pp}}\frac{\dot{E}_{\rm tide}}{\dot{E}_{\rm pp}} \diff f.
\end{align}
In the above derivation we have decomposed the total frequency evolution rate $\dot{f}$ as the sum of a point-particle (``pp'') part $\dot{f}_{\rm pp}$ driven by the GW radiation and a tidal contribution $\dot{f}_{\rm tide}$.  We have also treated the tidal effect as a small perturbation and have assumed that the orbit remains quasi-circular, which allows us to relate the tidally induced GW frequency shift to the excess energy dissipation as $\dot{f}_{\rm tide}/\dot{f}_{\rm pp}\simeq \dot{E}_{\rm tide}/\dot{E}_{\rm pp}$. Here $\dot{E}_{\rm tide} = \dot{E}_{\rm tide1} + \dot{E}_{\rm tide2}$ [cf.\ eq.~(\ref{eq:dEtide})].

Similarly, the time-domain phase can be written as 
\begin{align}
\phi\left[t(f)\right]&=2\pi \int \frac{f}{\dot{f}} \diff f \nonumber \\
                          &\simeq 2\pi \int \frac{f}{\dot{f}_{\rm pp}}\diff f - 2\pi \int\frac{f}{\dot{f}_{\rm pp}}\frac{\dot{E}_{\rm tide}}{\dot{E}_{\rm pp}} \diff f \nonumber \\
                          &=\phi_{\rm pp}\left[t(f)\right]  - 2\pi \int\frac{f}{\dot{f}_{\rm pp}}\frac{\dot{E}_{\rm tide}}{\dot{E}_{\rm pp}} \diff f.
\end{align}

Thus the frequency-domain phase can now be written as 
\begin{align}
\Psi(f) &= \Psi_{\rm pp}(f) - 2\pi \left(f \int \frac{1}{\dot{f}_{\rm pp}}\frac{\dot{E}_{\rm tide}}{\dot{E}_{\rm pp}}\diff f - \int\frac{f}{\dot{f}_{\rm pp}}\frac{\dot{E}_{\rm tide}}{\dot{E}_{\rm pp}} \diff f \right).
\end{align}
The lower and upper limits of the integrals are $f_0$ and $f$, respectively, where $f_0$ is the initial frequency of the signal. Therefore we always align the tidal waveform to the point-particle one at the beginning of the signal.

\bibliography{references}

\end{document}

%% file: LocalizationTable.tex
HLV & $7.9\times 10^{-3}$ & $4.1\times 10^{-2}$ & $1.1\times 10^{-2}$ & $5.4\times 10^{-2}$ \\
HLVKA & $2.0\times 10^{-3}$ & $5.6\times 10^{-3}$ & $3.1\times 10^{-3}$ & $8.5\times 10^{-3}$ \\
T & $3.5\times 10^{-5}$ & $5.4\times 10^{-5}$ & $4.4\times 10^{-3}$ & $1.1\times 10^{-2}$ \\
HLVKA + L2 T & $1.6\times 10^{-5}$ & $2.9\times 10^{-5}$ & $5.4\times 10^{-4}$ & $1.5\times 10^{-3}$ \\
HLVKA + $5^\circ$ T & $5.7\times 10^{-6}$ & $1.3\times 10^{-5}$ & $6.6\times 10^{-5}$ & $1.9\times 10^{-4}$ \\
HLVKA + $20^\circ$ T & $1.3\times 10^{-6}$ & $3.5\times 10^{-6}$ & $1.6\times 10^{-5}$ & $5.2\times 10^{-5}$ \\